\documentclass[12pt]{article}
\usepackage[dvips]{graphicx}
\usepackage{amsmath}
\usepackage{amssymb}
\usepackage{amsfonts}
\usepackage{rotating}
\usepackage{overcite}
\usepackage{ccaption}
\usepackage{color}
\usepackage[colorlinks,urlcolor=black]{hyperref}

\begin{document}
\setcounter{page}{1}

\title{\textbf{High-throughput sequencing reveals a simple model of nucleosome energetics}}

\author{
George Locke$^\dag$, Denis Tolkunov$^\dag$, Zarmik Moqtaderi$^\ddag$, \\
Kevin Struhl$^\ddag$ and Alexandre V.~Morozov$^{\dag\ast}$ \\
{\footnotesize $^\dag$Department of Physics \& Astronomy  and BioMaPS Institute for Quantitative Biology, Rutgers University} \\
{\footnotesize Piscataway, NJ 08854} \\
{\footnotesize $^\ddag$Department of Biological Chemistry and Molecular Pharmacology, Harvard Medical School} \\
{\footnotesize Boston, MA 02115} \\
{\footnotesize $^\ast$Corresponding author; ~~email: morozov@physics.rutgers.edu; ~~phone: 732-445-1387; ~~fax: 732-445-5958}
}

\date{}
\maketitle

\begin{abstract}
We use nucleosome maps obtained by high-throughput sequencing to
study sequence specificity of intrinsic histone-DNA interactions.
In contrast with previous approaches, we employ an analogy between
a classical one-dimensional fluid of finite-size particles in an
arbitrary external potential and arrays of DNA-bound histone
octamers. We derive an analytical solution to infer free energies
of nucleosome formation directly from nucleosome occupancies
measured in high-throughput experiments. The sequence-specific
part of free energies is then captured by fitting them to a sum of
energies assigned to individual nucleotide motifs. We have
developed hierarchical models of increasing complexity and spatial
resolution, establishing that nucleosome occupancies can be
explained by systematic differences in mono- and dinucleotide
content between nucleosomal and linker DNA sequences, with
periodic dinucleotide distributions and longer sequence motifs
playing a secondary role. Furthermore, similar sequence signatures
are exhibited by control experiments in which genomic DNA is
either sonicated or digested with micrococcal nuclease in the
absence of nucleosomes, making it possible that current
predictions based on high-throughput nucleosome positioning maps
are biased by experimental artifacts.
\end{abstract}

\section{Introduction}

In eukaryotes, 75-90\% of genomic DNA is packaged into histone-DNA complexes called nucleosomes. \cite{vanholde:1989}
Each nucleosome consists of 147 base pairs (bp) of DNA wrapped around a histone octamer in a left-handed superhelix. \cite{richmond:2003}
Arrays of nucleosomes fold into filamentous chromatin fibers which constitute building blocks for higher-order
structures. \cite{felsenfeld:2003} DNA wrapped in a nucleosome is occluded from interacting with other DNA-binding proteins such as
transcription factors, RNA polymerase, and DNA repair complexes. \cite{richmond:2003} On the other hand, histone tail domains act as substrates for
post-translational modifications, providing binding sites for chromatin-associated proteins which facilitate transitions
between active and silent chromatin states. \cite{jenuwein:2001}

Several distinct factors affect nucleosome positions in living cells.
First of all, intrinsic histone-DNA interactions are sequence-specific: for example,
poly(dA:dT) tracts are well-known to disfavor nucleosome formation. \cite{kaplan:2009,sekinger:2005}
In addition, nucleosome-depleted regions can be generated through the action of ATP-dependent chromatin remodeling enzymes \cite{becker:2002} and histone
acetylases. \cite{deckert:2001} Finally, non-histone DNA-binding factors can alter nucleosome positions through binding their cognate sites and either displacing
nucleosomes or hindering their subsequent formation. \cite{adams:1995,miller:2003}

The nucleosome code hypothesis states that DNA sequence is the primary determinant of nucleosome positions in living cells. \cite{segal:2006}
This hypothesis is often contrasted with the idea of statistical positioning which asserts that most nucleosomes are ordered into regular arrays
simply by steric exclusion. \cite{kornberg:1988,fedor:1988} In this view the nucleosomal arrays are ``phased'' by external boundaries such as
DNA-bound factors or DNA sequences unfavorable for nucleosome formation. It is also possible that a small number of nucleosomes
with favorable binding affinities create boundaries against which neighboring nucleosomes are ordered
by steric exclusion. \cite{mavrich:2008,mavrich:2008b}

Nucleosome positioning can be thought of as rotational,
referring to the 10-11 bp-periodic orientation of the DNA helix with respect to the surface of the histone octamer, and translational, referring to
the 147 bp-long sequence covered by a particular histone octamer. 
Optimal rotational positioning minimizes free energy of sequence-specific DNA bending, causing 10-11 bp periodicity of dinucleotide frequencies
in nucleosome positioning sequences. \cite{widom:2001}
We use a probabilistic description of translational positioning in which 147 bp sites with favorable free energies of nucleosome formation
have a higher probability to be nucleosome-covered.

To study the contribution of intrinsic histone-DNA interactions to nucleosome positioning,
several computational models based solely on the DNA sequence have been developed.
These models can be divided into
bioinformatics, which are trained on sets of nucleosomal sequences obtained from living cells \cite{segal:2006,ioshikhes:2006,field:2008,yuan:2008,peckham:2007,lee:2007}
or from \textit{in vitro} reconstitution experiments \cite{kaplan:2009}, and
\textit{ab initio}, which predict nucleosome energies and occupancies using DNA elasticity theory and structural data. \cite{morozov:2009,miele:2008,tolstorukov:2007}

Here we develop a physical model for predicting free energies of nucleosome formation directly from high-throughput maps of nucleosome positions.
Our model employs an exact relation between measured nucleosome occupancies and free energies, treating steric exclusion rigorously
in the presence of histone-DNA interactions of arbitrary strength and sequence specificity.
We focus in particular on nucleosomes reconstituted \textit{in vitro} on yeast genomic DNA. \cite{zhang:2009}
In this case nucleosome locations are affected solely by intrinsic histone-DNA interactions and by formation of higher-order chromatin structures.
We compare our predictions with sequence signals from two nucleosome-free control experiments in which DNA was either sonicated or digested
with micrococcal nuclease (MNase) to yield mononucleosome-size segments. We also test the ability of our \textit{in vitro} model to predict nucleosome
positions \textit{in vivo} and study the universality of nucleosome positioning motifs by applying our approach to other organisms.


\section{Results}

\noindent \textbf{Biophysical model of nucleosome occupancy and energetics.}

We have predicted histone-DNA energies genome-wide using an analogy between arrays of nucleosomes and
a one-dimensional fluid of non-overlapping particles of size $147 ~\mathrm{bp}$ in an arbitrary external potential.
The nucleosomal array is a discrete version of a one-dimensional system of finite-size particles for which Jerry K. Percus showed
that particle energies can be inferred exactly from the density profile. \cite{percus:1976}
Although our method neglects formation of three-dimensional chromatin structures which may cause linker DNA
to adopt preferred lengths, \cite{ulanovsky:1986,wang:2008,lubliner:2009} it rigorously takes into account both steric exclusion between neighboring particles
and intrinsic histone-DNA interactions, including the 10-11 bp periodic rotational component.
Our approach, outlined in Fig. \ref{fig_locus},
proceeds in the direction opposite to previous work which first employed either bioinformatics or DNA elastic theory to construct a sequence-specific
histone-DNA interaction potential and then positioned nucleosomes on genomic DNA without steric overlap. \cite{segal:2006,field:2008,kaplan:2009,morozov:2009}
In contrast, we employ an exact decomposition from experimentally available nucleosome probabilities and occupancies to free energies of
nucleosome formation which we call Percus energies (Eq. \ref{MTeq:1}).

To extract the sequence-specific component of nucleosome energetics, we fit Percus energies at each genomic bp to a
sum of energies of individual nucleotide motifs
ranging from $1$ to $N$ bp in length (see Methods).
There is no need to construct an explicit background model of word frequencies with this approach.
The words with the same nucleotide sequence have the same energy if they occur anywhere in the 147 bp-long nucleosomal site
(the position-independent model, Eq. \ref{MTeq:2}), or fall into one of the three equal-length regions that span the 147 bp site (the three-region model),
or are separated by an integer multiple of the 10 bp DNA helical twist (the periodic model).
All models are constrained to assign non-zero energies to words
with $N$ nucleotides only if the sequence specificity of Percus energies cannot be captured
using words with $1 \dots N-1$ nucleotides (see Supplementary Methods).
We refer to the maximum length of the words included into a model as its \emph{order} $N$.
In addition, we have developed an order 2 model in which mono- and dinucleotides are allowed to have different energies at every position
in the 147 bp-long nucleosomal site (the spatially resolved model, Eq. \ref{MTeq:3}). 

The sequence-specific models provide nucleosome formation energies at each bp in \textit{S.cerevisiae}, \textit{C.elegans} and \textit{E.coli}
genomes. These energies serve as input to a standard recursive algorithm which computes probabilities to start a nucleosome
at each genomic bp and thus nucleosome occupancies (defined as the probability that a given bp is covered
by any nucleosome, see Methods). \cite{segal:2006,morozov:2009}

\noindent \textbf{A:T/G:C content is the primary determinant of nucleosome sequence preferences in \textit{S.cerevisiae}.}

\begin{figure}[tbph]
\centering
\includegraphics[scale=0.80]{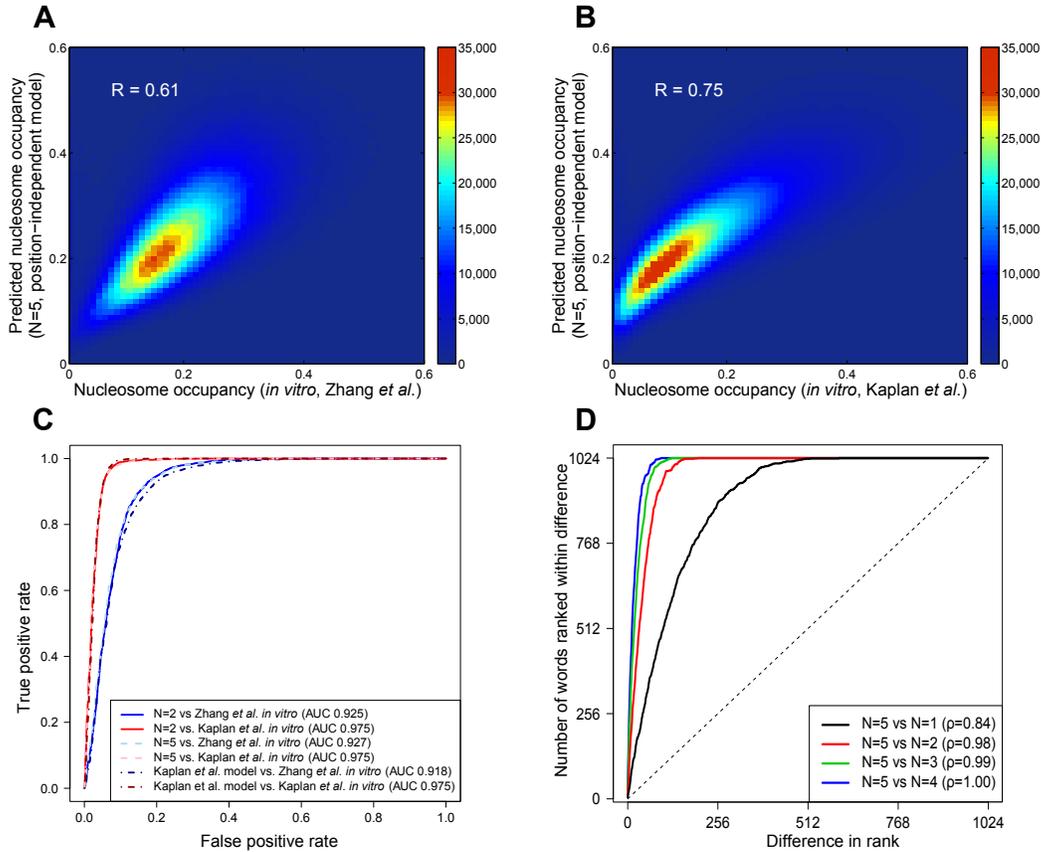}
\caption{ \label{fig_N2_OK}
\textbf{Position-independent model predicts \textit{in vitro} nucleosome occupancy in \textit{S.cerevisiae}
with high accuracy.}
a) Density scatter plot for the nucleosome occupancy at each genomic base pair predicted with the $N=5$ position-independent model
vs. \textit{in vitro} occupancy observed by Zhang \textit{et al.} \cite{zhang:2009}
The color of each region represents the number of data points mapped to that region. Our model is fit on this data (see Methods).
b) Same as (a) except the \textit{in vitro} occupancy is from Kaplan \textit{et al.} \cite{kaplan:2009}
c) The receiver operating characteristic (ROC) curve for discriminating between DNA segments with high and low nucleosome occupancy.
The yeast genome was parsed into 500 bp windows and the average nucleosome occupancy was computed for each window.
5000 windows with the highest and 5000 with the lowest average occupancies were ranked high-to-low using occupancies predicted
with the $N=2$ position-independent model, $N=5$ position-independent model, and Kaplan \textit{et al.} model. \cite{kaplan:2009}
For each partial list of ranked windows with $1, \dots, 10000$ entries we plot the number of windows in the list
known to have high occupancy on the y-axis, low occupancy on the x-axis.
d) Rank-order plots of energies of 5 bp words: the energy of each word is ranked using position-independent models of order $N=1$ through $N=4$ and
compared with the $N=5$ model. Each curve shows the number of words whose ranks are separated by a given distance or less.
}
\end{figure}

The $N=5$ position-independent model, which assigns energies to 364 independent words from 1 bp to 5 bp in length, is capable of accurately predicting
occupancy by nucleosomes assembled \textit{in vitro} on yeast genomic DNA (Fig. \ref{fig_N2_OK}a,b). Remarkably,  even though the model is based on
Zhang \textit{et al.} high-throughput nucleosome positioning data (yielding r=0.54 on average between Percus energies and sequence-specific energies
fit independently on 10 DNA segments of equal length spanning the yeast genome,
and r=0.61 between predicted and observed occupancies), \cite{zhang:2009}
its prediction of Kaplan \textit{et al.} \textit{in vitro} occupancies \cite{kaplan:2009}
is more accurate (r=0.75), partially due to the 2.85-fold higher sequence coverage in the latter dataset.
Indeed, the correlation coefficient drops from 0.75 to 0.70 when sequence reads are randomly removed from the Kaplan \textit{et al.} map
to match Zhang \textit{et al.} level of read coverage.

The correlation between the two \textit{in vitro} datasets is rather low (r=0.69), probably
because Kaplan \textit{et al.} assembled \textit{in vitro} chromatin at less than physiological histone
octamer concentrations. \cite{kaplan:2009}
The $N=5$ model is also highly successful in discriminating between high- and low-occupancy regions (dashed curves in Fig. \ref{fig_N2_OK}c).
Its performance is comparable to the Kaplan \textit{et al.} bioinformatics model \cite{kaplan:2009} which 
takes both distributions of 5 bp-long words in nucleosomes and linkers and position-dependent
dinucleotide frequencies into account (Supplementary Table \ref{ST1}, dotted curves in Fig. \ref{fig_N2_OK}c).
Occupancies predicted by the two models are highly correlated (r=0.89) and thus capture essentially the same nucleosome sequence preferences.
Note that we report correlations between occupancy profiles while Kaplan \textit{et al.} log-transform occupancies before
computing a linear correlation coefficient: as a result we obtain r=0.79 between Kaplan \textit{et al.} predicted and \textit{in vitro} occupancies,
whereas they report r=0.89 for the same comparison. \cite{kaplan:2009}

However, we find that using 5 bp-long words is not necessary: $N=2 \dots 4$ position-independent models are virtually identical to the $N=5$ model
in ranking 5 bp-long sequences (Fig. \ref{fig_N2_OK}d), classifying high- and low-occupancy regions (solid curves in Fig. \ref{fig_N2_OK}c), and predicting \textit{in vitro}
nucleosome occupancies (Supplementary Fig. \ref{Sfig_N2}a,b). The $N=2$ model remains highly correlated with the Kaplan \textit{et al.} bioinformatics model
(r=0.89; Supplementary Fig. \ref{Sfig_Kaplan_model}). Remarkably, even the $N=1$ position-independent model with one free parameter
($\epsilon_A = \epsilon_T$ and $\epsilon_C = \epsilon_G$ if both DNA strands are included for each mapped nucleosome) retains
most of the predictive power of the higher-order models (Fig. \ref{fig_N2_OK}d, Supplementary Table \ref{ST1}), in agreement with a recent
independent study. \cite{tillo:2009} Thus positions of nucleosomes reconstituted \textit{in vitro} on the yeast genome
are largely controlled by the differences in frequencies of A:T and G:C dinucleotides in nucleosomes and linkers.
In particular, higher-order terms play little role in the energetics of poly(dA:dT) tracts (Supplementary Fig. \ref{Sfig_polyA}).

\begin{figure}[tbph]
\centering
\includegraphics[scale=0.65]{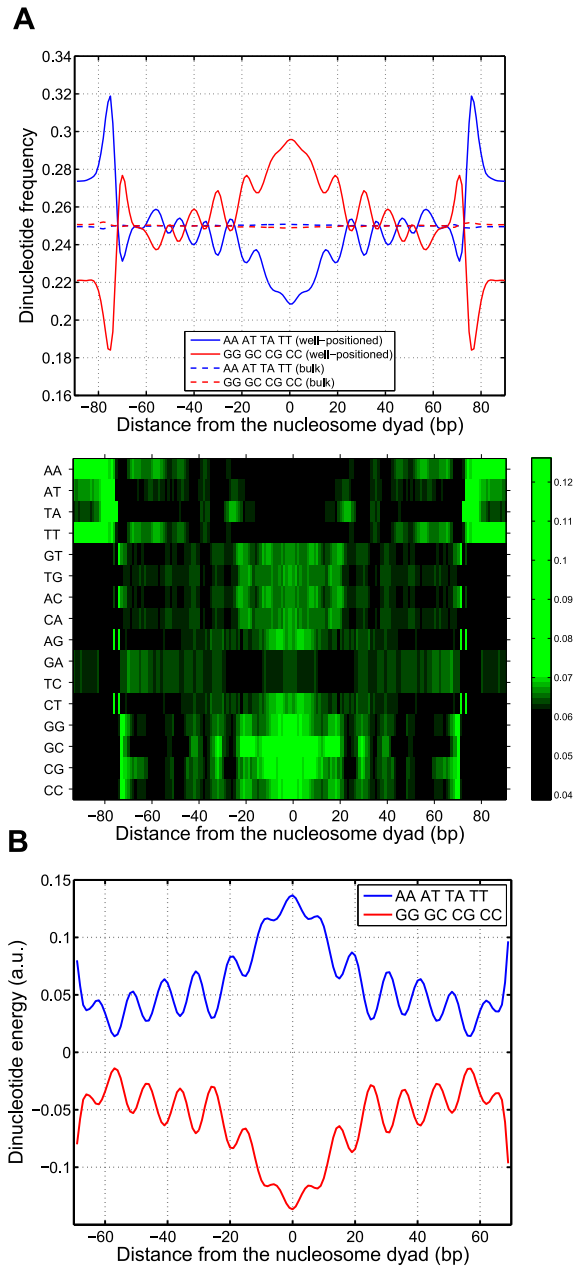}
\caption{ \label{fig_spatially_resolved}
\textbf{Dinucleotide distributions in nucleosome and linker sequences.}
a) \emph{Upper panel}: average relative frequencies of WW (AA, TT, AT and TA) and SS (CC, GG, CG and GC) dinucleotides at each position within the nucleosome are plotted
with respect to the nucleosome dyad. The relative frequency of each dinucleotide is defined as its frequency at a given position divided by genome-wide frequency.
All frequencies are smoothed using a 3 bp moving average.
Solid lines: well-positioned nucleosomes defined by five or more sequence reads, dashed lines: bulk nucleosomes defined by one or two sequence reads.
\emph{Lower panel}: heat map of relative frequencies for each dinucleotide, plotted with respect to the nucleosome dyad.
Nucleosomes were assembled \textit{in vitro} on the yeast genome using salt dialysis. \cite{zhang:2009}
b) Average energies of WW (AA, TT, AT and TA) and SS (CC, GG, CG and GC) dinucleotides at each position within the nucleosome predicted with
the $N=2$ spatially resolved model are plotted with respect to the nucleosome dyad.
}
\end{figure}

Indeed, Fig. \ref{fig_spatially_resolved}a shows that DNA sequences of well-positioned nucleosomes (defined by 5 or more sequence reads mapped
to the same genomic coordinates) are characterized by 
sharp A:T/G:C discontinuities across the nucleosome boundary. Overall, A:T nucleotides are depleted
in nucleosomes and enriched in linkers, with the opposite true for G:C nucleotides. Although well-positioned nucleosomes
make up only 5.4 \% of all mapped nucleosomes defined by one or more sequence reads, they produce an occupancy profile which
is highly correlated with the total nucleosome occupancy (r=0.71, with 56.4 \% of genomic bps covered by at least one well-positioned
nucleosome).
In contrast, 81.5 \% of all nucleosomes are defined by just one or two reads and
exhibit little sequence specificity (dashed lines in Fig. \ref{fig_spatially_resolved}a).
Furthermore, the $N=2$ position-independent model based only on well-positioned nucleosomes
is virtually identical to the $N=2$ model based on all nucleosomes (rank correlation of 0.94 between the two sets of dinucleotide energies).
Thus our predictions reveal sequence preferences of a subset of nucleosomes that tend to occupy unique sites on the DNA.

\noindent \textbf{Periodic motif distributions do not play a significant role in nucleosome occupancy predictions.}

Besides the A:T/G:C discontinuities, Fig. \ref{fig_spatially_resolved}a reveals two additional features that could affect positioning preferences of yeast
nucleosomes: prominent 10-11 bp dinucleotide periodicity and a particularly strong A:T depletion and G:C enrichment within 20 bp of the nucleosome dyad.
To test the utility of these features in nucleosome occupancy predictions we have employed three additional models
that either partially or fully differentiate between words located at different positions within the nucleosomal site.

The three-region model assigns different
energies to words found in the 47 bp-long core and flanking regions and is thus capable of capturing prominent dinucleotide biases in
the vicinity of the nucleosomal dyad, the periodic model enforces 10 bp helical twist periodicity while disregarding global effects, while the most detailed spatially
resolved model mirrors all three main features exhibited by the frequencies of dinucleotides
found in nucleosome positioning sequences (Fig. \ref{fig_spatially_resolved}b). Nevertheless, these models do not offer a significant improvement
over the $N=2$ position-independent model (Table \ref{T1}, Supplementary Table \ref{ST1}),
reflecting the fact that all three features are simultaneously present in well-positioned \textit{in vitro} nucleosomes (Fig. \ref{fig_spatially_resolved}a)
and so knowing any one of them is sufficient.
Furthermore, global A:T/G:C discontinuities appear to play the role of the primary nucleosome positioning determinant,
whereas periodic dinucleotide distributions and local enrichments can be greatly diminished or absent in other organisms and in \textit{in vivo} nucleosome
positioning maps from yeast (Supplementary Fig. \ref{Sfig_green_blue_red}).

However, the rotational positioning component of the yeast model should be more predictive for nucleosomes positioned on DNA sequences
with prominent 10-11 bp dinucleotide periodicities.
Indeed, the spatially resolved model works better than
the $N=2$ position-independent model for six non-yeast nucleosomes whose \textit{in vitro} positions on short ($< 250$ bp) DNA sequences
have been determined with single bp precision by hydroxyl radical footprinting (Supplementary Fig. \ref{Sfig_6nucs}).

\noindent \textbf{\textit{In vivo} nucleosome positions are partially controlled by the underlying DNA sequence.}

We investigated whether simple rules that govern \textit{in vitro} nucleosome positions remain valid in living cells
where chromatin structure may be affected by remodeling enzymes and by competition with non-histone DNA-binding factors. Indeed,
\textit{in vivo} nucleosomes appear to be well-positioned in the vicinity of transcription start and termination sites, with prominent nucleosome-depleted regions
(NDRs) on both ends of the transcript (Supplementary Fig. \ref{Sfig_TSS_TTS}; \textit{in vivo} chromatin comes from cells grown in YPD medium \cite{kaplan:2009}).
In contrast, \textit{in vitro} nucleosomes are much more delocalized (Supplementary Fig. \ref{Sfig_disthist}),
so that nucleosomal arrays around NDRs are not ordered and NDRs themselves are much less pronounced (Supplementary Fig. \ref{Sfig_TSS_TTS}). \cite{zhang:2009}

Despite these differences, dinucleotide energies extracted from \textit{in vitro} and \textit{in vivo} nucleosome positioning maps are reasonably well correlated
(Supplementary Fig. \ref{Sfig_invivo}), yielding nearly identical predictions of Zhang \textit{et al.} \textit{in vitro} nucleosome occupancies (Supplementary Table \ref{ST1}).
Although dinucleotide energies inferred from the \textit{in vivo} map of cross-linked nucleosomes
are not as close to their \textit{in vitro} counterparts as the energies based on the \textit{in vivo} map without cross-linking,
the two \textit{in vivo} models yield very similar occupancy profiles (r=0.94).
This indicates that MNase binding does not significantly alter nucleosome positions.
Clearly, the striking oscillations observed in the \textit{in vivo} occupancy profile (Supplementary Fig. \ref{Sfig_TSS_TTS}) are not due to intrinsic sequence preferences
but rather involve biological factors such as components of transcription initiation
machinery that may act to position the first nucleosome downstream of the NDR (the so-called +1 nucleosome \cite{mavrich:2008}). \cite{zhang:2009}

\noindent \textbf{Energetics of nucleosome formation in \textit{E.coli} and \textit{C.elegans}.}

\begin{figure}[tbph]
\centering
\includegraphics[scale=0.92]{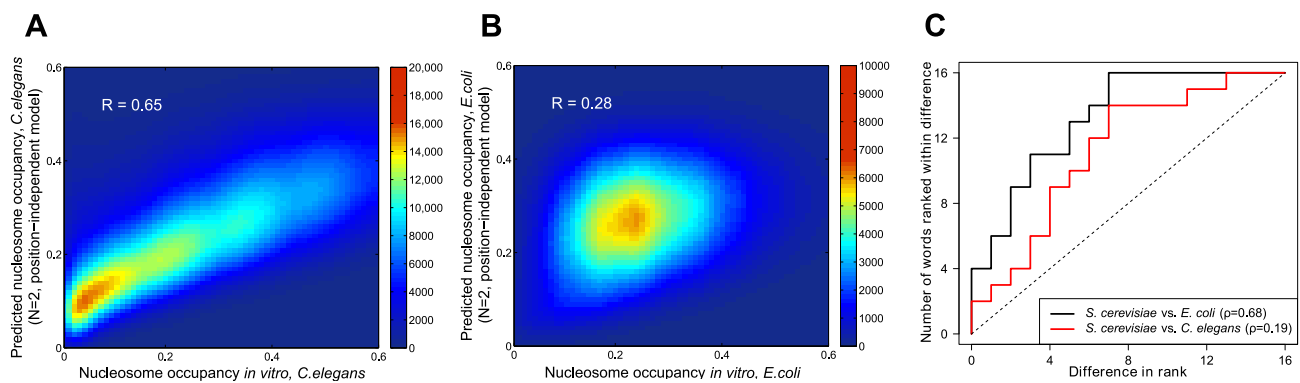}
\caption{ \label{fig_organisms}
\textbf{Prediction of nucleosome occupancies in \textit{C.elegans} and \textit{E.coli}.}
Density scatter plots for the nucleosome occupancy at each genomic base pair (predicted with the $N=2$ position-independent model)
vs. \textit{in vivo} occupancy in \textit{C.elegans} \cite{valouev:2008} (a) and  \textit{in vitro} occupancy in \textit{E.coli} \cite{zhang:2009} (b).
Rank-order plots of energies of 2 bp words (c): the energy of each word is ranked using a position-independent model of order $N=2$ trained on
either \textit{in vitro} (\textit{S.cerevisiae}, \textit{E.coli}) or \textit{in vivo} (\textit{C.elegans}) nucleosome
positioning data. Each curve shows the number of words whose ranks are separated by a given distance or less in the \textit{C.elegans} and \textit{E.coli}
vs. \textit{S.cerevisiae} fits.
}
\end{figure}

To study whether dinucleotide-based nucleosome positioning patterns observed in \textit{S.cerevisiae} extend to other organisms,
we have inferred position-independent
dinucleotide energies from a map of nucleosomes assembled \textit{in vitro} on the \textit{E.coli} genome. \cite{zhang:2009}
Although the correlation between observed and predicted occupancies was modest in this case (Fig. \ref{fig_organisms}b),
probably because the \textit{E.coli} genome did not evolve to favor nucleosome formation
(resulting in lower sequence read coverage in competition with yeast DNA),
the dinucleotide energies were similar in yeast and \textit{E.coli} (Fig. \ref{fig_organisms}c).
The most prominent difference was exhibited by the four CG-containing dinucleotides which have the lowest energies in \textit{S.cerevisiae}
but occupy middle positions in the case of \textit{E.coli} (Supplementary Table \ref{ST2}).

Dinucleotide energies inferred from the \textit{in vivo} map of \textit{C.elegans} nucleosomes, \cite{valouev:2008} while an excellent predictor of
nucleosome occupancies in the \textit{C.elegans} genome (Fig. \ref{fig_organisms}a), are even further from their yeast counterparts (Fig. \ref{fig_organisms}c):
$\epsilon_{CG}$ and $\epsilon_{GC}$ become comparable to $\epsilon_{AA/TT}$ (which is the highest in all three organisms), whereas
$\epsilon_{CC/GG}$ is close to $\epsilon_{TA}$ which is the third most unfavorable in yeast (Supplementary Table \ref{ST2}).
It is possible that \textit{in vivo} effects override intrinsic nucleosome preferences in \textit{C.elegans}.
In addition, we find that the mononucleotide model is much less predictive in this organism:
$N=2$ and $N=1$ position-independent models yield r=0.65 and r=0.45 correlations with the \textit{in vivo} map to which they were fit,
vs. r=0.60 and r=0.54 for the same models applied to \textit{S.cerevisiae} \textit{in vitro} nucleosomes (Supplementary Table \ref{ST1}).
On the other hand, fitting energies of 3 bp-long words resulted only in a slight (3.0 \%) improvement in the correlation coefficient,
indicating that it is not necessary to keep track of higher-order motifs in \textit{C.elegans}.

Although the dinucleotide energies are somewhat different in the three organisms we examined, position-independent models
from one organism can still be used to predict nucleosome positions in another. For example, using the $N=2$ \textit{E.coli} model to predict
\textit{in vitro} nucleosome occupancies in \textit{S.cerevisiae} \cite{zhang:2009} results in r=0.55, which is only a little worse than r=0.60 observed
with the ``native'' model (Supplementary Table \ref{ST1}).
The $N=2$ \textit{C.elegans} model has a correlation of 0.46 with the \textit{in vitro} occupancy from \textit{S.cerevisiae},
while the correlation between the $N=2$ \textit{S.cerevisiae} \textit{in vitro} model and the \textit{in vivo} occupancy from \textit{C.elegans} is 0.52, somewhat lower than
0.65 obtained with the ``native'' model.
Therefore it is possible to make useful predictions in organisms for which
high-throughput nucleosome positioning maps are not yet available.

\noindent \textbf{Nucleosome-free control experiments can be used to predict nucleosome positions.}

Depletion of A:T and enrichment of G:C-containing dinucleotides in nucleosomal sequences and the discontinuity of dinucleotide frequencies
across the nucleosome boundary may be an experimental artifact caused by MNase sequence specificity \cite{horz:1981} rather than a reflection of intrinsic histone-DNA interactions.
To study this possibility we have examined two collections of sequence reads obtained from nucleosome-free control experiments.
In one experiment a mixture of genomic DNA from \textit{S.cerevisiae} and \textit{E.coli} was digested with MNase, gel-purified to isolate $\sim 150$ bp DNA segments,
and sequenced.
In the other experiment DNA was sonicated rather than MNase-digested (see Methods). Because
DNA segments are approximately constant in length we can compute Percus energies and analyze their sequence specificity
using $N=2$ position-independent models (Fig. \ref{fig_locus}).

\begin{figure}[tbph]
\centering
\includegraphics[scale=0.35]{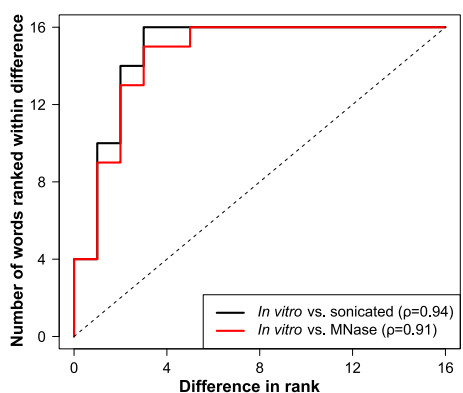}
\caption{ \label{fig_controls}
\textbf{Nucleosome-free control experiments yield sequences with nucleosome-like dinucleotide distributions.}
Rank-order plots of energies of 2 bp words: the energy of each word is ranked using a position-independent model of order $N=2$ trained on
either \textit{in vitro} nucleosome positioning sequences or fragments of mononucleosomal
size obtained from sonication and MNase digestion assays of nucleosome-free yeast DNA. \cite{zhang:2009}
Each curve shows the number of words whose ranks are separated by a given distance or less in the sonication
and MNase digestion vs. nucleosomal fits.
}
\end{figure}

Surprisingly, both control experiments yield dinucleotide energies that are very close to those obtained from the \textit{in vitro} nucleosome positioning map (Fig. \ref{fig_controls}).
The differences are smaller than those between the $N=2$ \textit{in vitro} model and the \textit{in vivo} model based on cross-linked nucleosomes
(Supplementary Fig. \ref{Sfig_invivo}), and are comparable to the discrepancies between two $N=2$ \textit{in vitro} models inferred from
Kaplan \textit{et al.} and Zhang \textit{et al.} datasets \cite{kaplan:2009,zhang:2009}
(which yield $\rho=0.96$ rank-order correlation coefficient between the two sets of dinucleotide energies). 
As a result, $N=2$ models trained on sonication and MNase controls predict Kaplan \textit{et al.} \textit{in vitro} nucleosome occupancies with correlation
coefficients of 0.64 and 0.58, respectively, compared with r=0.75 for the nucleosome model (Supplementary Table \ref{ST1}).

The most obvious explanation for this finding is that dinucleotide energies in nucleosomes reflect experimental artifacts.
Indeed, the distribution of dinucleotides frequencies in MNase-digested DNA segments of mononucleosome size is rather
similar to that observed in nucleosomes, especially in the vicinity of the segment boundary (Supplementary Fig. \ref{Sfig_green_blue_red}e).
Alternatively, nucleosome-free controls may be enriched in sequences that resemble nucleosome positioning sequences.
For example, sonication tends to break DNA segments across the A:T/G:C ``fault lines'', although the resulting
depletion of A:T and enrichment of G:C-containing dinucleotides are rather small (Supplementary Fig. \ref{Sfig_green_blue_red}f).
The degree to which currently available nucleosome maps are affected by experimental artifacts requires further studies
which do not rely on MNase digestion or sonication to isolate mononucleosome cores.

\section{Discussion}


Nucleosome positioning has been extensively studied using MNase
digestion to isolate mononucleosomal DNA, followed by either
microarray hybridization \cite{yuan:2005,lee:2007,zawadzki:2009}
or high-through-put sequencing.
\cite{valouev:2008,mavrich:2008,mavrich:2008b,field:2008,kaplan:2009,zhang:2009}
Several bioinformatics models have been fit to these data in order
to determine intrinsic histone-DNA sequence specificity and its
contribution to \textit{in vivo} chromatin structure. Until
recently, these models tended to be quite complex, taking both
periodic dinucleotide distributions and relative frequences of
longer motifs into account,
\cite{yuan:2008,field:2008,kaplan:2009} although it has also been
stated that simple models based on A:T/G:C content and related
descriptors are sufficient for predicting nucleosome occupancies.
\cite{peckham:2007,tillo:2009} However, these findings may be
subject to MNase and sequencing biases which inevitably affect
currently available nucleosome maps. Besides, bioinformatics
models are not capable of directly predicting nucleosome formation
energies.

To study these issues, we have developed a biophysical approach to inferring nucleosome energies and occupancies from high-throughput sequencing data.
The effects of steric exclusion is rigorously separated from intrinsic histone-DNA interactions under the assumption that nucleosomes form
a one-dimensional array in which there are no nucleosome-nucleosome interactions besides nearest-neighbor steric hindrance.
This assumption amounts to neglecting intrinsic structure of the chromatin fiber which is believed
to impose ``quantized'' linker lengths. \cite{ulanovsky:1986,widom:1992,wang:2008}
Furthermore, we assume that the one-dimensional nucleosome array is in thermodynamic equilibrium, with
individual nucleosome positions corresponding to the lowest free energy state of the entire array.
\textit{In vivo} nucleosomes may not be in equilibrium due to the action of
chromatin remodeling enzymes and other energy-dependent processes.

We find that most mapped nucleosomes are not sequence-specific. However, well-positioned nucleosomes defined by five or more
sequence reads tend to occupy G:C-enriched and A:T-depleted DNA segments in \textit{S.cerevisiae} (Fig. \ref{fig_spatially_resolved}a).
These nucleosomes alone 
define an occupancy profile which is highly correlated (r=0.71) with the profile based on all mapped nucleosomes.
Thus A:T and G:C-containing dinucleotide content is different in nucleosomal and linker sequences and is highly predictive of nucleosome positions.
More complex models that take rotational positioning into account do not yield siginificantly improved predictions (Table \ref{T1}), indicating that
10 bp dinucleotide periodicites alone do not define nucleosome positions on the yeast genome. 
It is possible that nucleosomal sequences first evolved to be G:C-rich and then acquired 10 bp dinucleotide periodicity
to take advantage of the rotational positioning mechanism.

Surprisingly, models trained on DNA segments from nucleosome-free control experiments can be used to predict nucleosome occupancies
(Fig. \ref{fig_controls},  Supplementary Table \ref{ST1}).
It is possible that experimental biases obscure nucleosome positioning signals in current high-throughput experiments. Alternatively, DNA from
control experiments may be enriched in nucleosome positioning sequences because A:T/G:C discontinuities make it easier
for sonication or MNase to break DNA at the nucleosome boundary even in the absence of nucleosomes.
MNase- and sonication-free nucleosome positioning maps are required to resolve this issue.

In summary,
nucleosome sequence preferences can be captured using a simple physical model based on dinucleotide content.
Promoter regions are unfavorable for nucleosome formation, while +1 nucleosomes have lower energies and help
define nucleosome array boundaries, although sequence preferences alone are not strong enough to explain perfectly phased \textit{in vivo} arrays
(Supplementary Figure \ref{Sfig_TSS_TTS}).
Similar nucleosome-positioning rules can be extracted from \textit{in vitro} and \textit{in vivo} chromatin (Supplementary Figure \ref{Sfig_invivo}),
strongly suggesting that nucleosomes tend to occupy thermodynamically favorable positions in living cells. \cite{kaplan:2009}

\section{Materials and Methods}

\noindent \textbf{Parallel sequencing and mapping of \textit{in vitro} and \textit{in vivo} nucleosomes.}

Nucleosomes were reconstituted \textit{in vitro} on \textit{S.cerevisiae} and \textit{E.coli} genomic
DNA as follows: \cite{zhang:2009}
genomic DNA from \textit{S.cerevisiae} and \textit{E.coli} was purified using Qiagen genomic tip 500/G.
Yeast and bacterial DNA were mixed at a 3:1 mass ratio.
5-10 kb DNA segments (obtained by light sonication of purified genomic DNA) were assembled into chromatin by salt dialysis
and extensively digested with MNase to yield mononucleosome core particles. Mononucleosomal DNA was purified  by excision from an agarose gel
and sequenced using Illumina Genome Analyzer. This procedure resulted in a collection of 3239990 25 bp-long sequence reads (0.27 reads per bp) mapped
to the \textit{S.cerevisiae} genome (SGD April 2008 build) and 336338 reads (0.07 reads per bp) mapped to the \textit{E.coli} K12 genome (U00096),
allowing up to 2 mismatches per read.


In control experiments, mixture of genomic DNA from \textit{S.cerevisiae} and \textit{E.coli} was prepared as described above.
Part of this mixture was treated with MNase (USB) to yield a small
average fragment size ($<300$ bp), and DNA fragments of approximately 150 bp were
purified by excision from an agarose gel.  A second fraction of the
yeast/bacterial DNA mixture was subjected to sonication in a Misonix
water-bath instrument to yield an average fragment size of 150bp.
Mononucleosome-size DNA fragments were sequenced using Illumina Genome Analyzer,
yielding 1160528 reads mapped to yeast (0.10 reads per bp) for the MNase-digested fraction
and 1326882 reads mapped to yeast (0.11 reads per bp) for the sonicated fraction.

We have also used maps of \textit{in vivo} nucleosomes prepared from log-phase yeast cells grown in YPD medium. \cite{kaplan:2009}
In two replicates nucleosomes were cross-linked with formaldehyde prior to MNase digestion, and in four replicates the cross-linking
step was omitted. We have combined sequence reads in each case, resulting in
0.50 and 1.50 mapped reads per bp respectively.
Kaplan \textit{et al.} also provide two replicates for nucleosomes reconstituted \textit{in vitro} on
yeast genomic DNA, with the total of 0.77 mapped reads per bp. \cite{kaplan:2009}
\textit{C.elegans} nucleosomes came from mixed stage, wild-type (N2) cells. \cite{valouev:2008}
Mononucleosome cores were liberated with MNase and sequenced on a SOLiD Analyzer (Applied Biosystems).
We have used sequence read coordinates provided by the authors, yielding 0.44 reads per bp.

\noindent \textbf{Pre-processing of nucleosome sequence reads.}

\begin{figure}[tbph]
\centering
\includegraphics[scale=0.58]{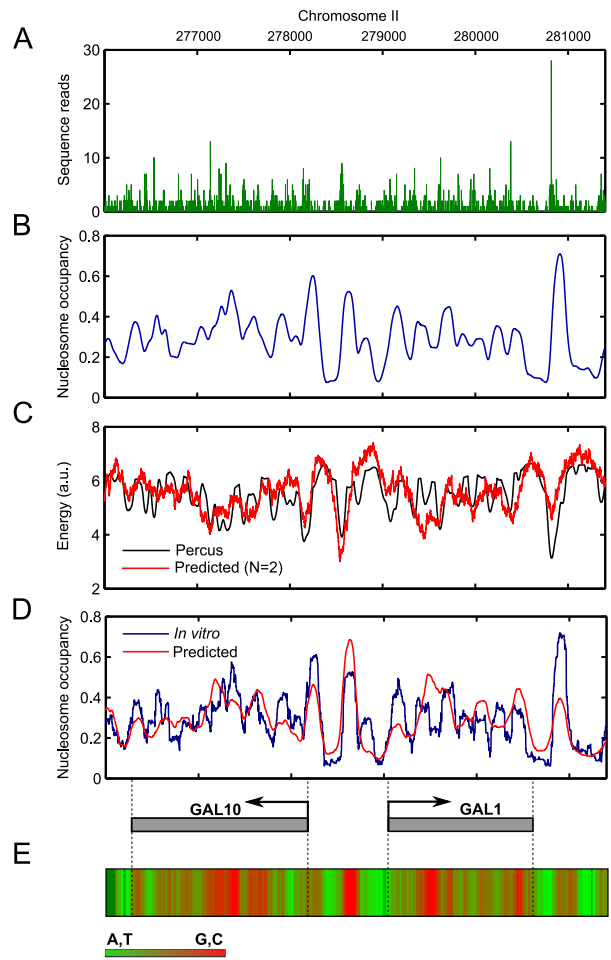}
\caption{ \label{fig_locus}
\textbf{Outline of the biophysical approach to nucleosome occupancy predictions:
\textit{GAL1-10} \textit{S. cerevisiae} locus.}
a) Nucleosome starting positions mapped to the \textit{GAL1-10} locus in the \textit{in vitro} reconstitution experiment. \cite{zhang:2009}
b) Nucleosome occupancy based on the nucleosome starting positions shown in (a) and smoothed with
a $\sigma=20$ Gaussian (see Supplementary Methods).
c) Percus energy inferred from the occupancy profile shown in (b), and a sequence-specific linear model fit to an $N=2$ position-independent model.
d) Nucleosome occupancy predicted using sequence-specific energies and compared with the experimental occupancy based on the
nucleosome starting positions shown in (a) (same as (b) but without Gaussian smoothing).
e) Nucleosomes are positioned over G:C-rich sequences:
shown are nucleotide counts in the \textit{GAL1-10} locus, 
smoothed with a 100 bp moving average.
}
\end{figure}

We assume that genomic coordinates of 25 bp-long mapped sequence reads define nucleosome positions on yeast and \textit{E.coli} genomes.
We extend all mapped reads to the 147 bp canonical nucleosome length and combine reads from both strands (Supplementary Methods).
This procedure yields the number of nucleosomes that start at each genomic bp (the sequence read profile; Fig. \ref{fig_locus}a), as well as the
number of nucleosomes that cover a given bp (the nucleosome coverage profile).
We control for sequencing and mapping artifacts by removing regions with anomalously
high and low nucleosome coverage from further consideration (Supplementary Methods).

Next we smooth the sequence read and nucleosome coverage profiles by replacing the number of nucleosomes starting
at each bp with a Gaussian centered on that bp (Fig. \ref{fig_locus}b). \cite{mavrich:2008,mavrich:2008b} The area of the Gaussian is equal to the number
of sequence reads starting at that position, while its $\sigma$ is set to either 2 or 20.
Gaussian smoothing is necessary because current levels of sequence read coverage lead to large deviations in the number of nucleosomes
located at neighboring bps, contrary to the expectation that such nucleosomes should have very similar binding affinities because they
occupy nearly identical sites. \cite{segal:2006} The effect of Gaussian smoothing can be seen in Supplementary Fig. \ref{Sfig_acf}.

Finally, we normalize the sequence read and nucleosome coverage profiles by the
highest value of nucleosome coverage on the chromosome.
We interpret the resulting normalized profiles as the probability to start a nucleosome at a given bp (the nucleosome probability profile)
and the probability that a given bp is covered by any nucleosome (the nucleosome occupancy profile).

\noindent \textbf{Prediction of nucleosome energetics from high-throughput sequencing maps.}

We derive nucleosome formation energies directly from the smoothed probability and occupancy profiles, under the assumption that
observed nucleosome positions are affected solely by intrinsic histone-DNA interactions and steric exclusion (Supplementary Methods):

\begin{equation} \label{MTeq:1}
\frac{E_i - \mu}{k_B T} = \log \frac{1 - O_i + P_i}{P_i} + \sum_{j=i}^{i+146} \log \frac{1 - O_j}{1 - O_j + P_j}, \quad i = 1, \dots, L-146
\end{equation}

Here $E_i$ is the Percus energy at bp $i$, $\mu$ is the chemical potential of histone octamers,
$k_B T$ is the product of the Boltzmann constant and room temperature,
$L$ is the number of bps in the DNA segment, $P_i$ is the probability to start a nucleosome at bp $i$,
and $O_i$ is the nucleosome occupancy of bp $i$ ($O_i = \sum_{j = i-146}^{i} P_j$).

We establish the degree of correlation between Percus energies and sequence features found in nucleosomal and linker DNA
by fitting them to four sequence-specific models (Fig. \ref{fig_locus}c).
The position-independent model of order $N$ is given by:

\begin{equation} \label{MTeq:2}
\frac{E_i - \mu}{k_B T} = \sum_{n = 1}^{N} \sum_{\{\alpha_{1} \dots \alpha_{n}\}}^{3^n} n_{\alpha_{1} \dots \alpha_{n}}^{i} \epsilon_{\alpha_{1} \dots \alpha_{n}} + \epsilon^0,
\end{equation}
where $N$ is the maximum word length, $\epsilon^0$ is the total offset, and $n_{\alpha_{1} \dots \alpha_{n}}^{i}$ is the number of times
a word of length $n$ with sequence $\alpha_1 \dots \alpha_n$ was found within the nucleosome
that started at bp $i$. $\epsilon_{\alpha_{1} \dots \alpha_{n}}$ are
word energies which do not depend on the nucleosome position $i$.
The word energies are constrained by $\sum_{\alpha_i} \epsilon_{\alpha_1 \dots \alpha_n} = 0, ~~\forall i = 1 \dots n$,
which leaves $3^n$ independent words of length $n$. We exclude all words that extend into 3 terminal bps on each end of the 147 bp
nucleosomal site from our counts.

The spatially resolved model is defined by:

\begin{equation} \label{MTeq:3}
\frac{E_i - \mu}{k_B T} = \sum_{j=i+3}^{i+143} \epsilon_{\alpha_j \alpha_{j+1}} +  \sum_{j=i+3}^{i+144} \epsilon_{\alpha_j} + \epsilon^0,
\end{equation}
where the mono- and dinucleotide energies are constrained as above at each position within the nucleosomal site.
The three-region model and the periodic model are described in Supplementary Methods.
We use Gaussian smoothing with $\sigma = 20$ for position-independent and three-region models and $\sigma = 2$ for spatially resolved
and periodic models.



Eqs. (\ref{MTeq:2}) and (\ref{MTeq:3}) define linear models which we fit against Percus energies using function \textit{lm} from R statistical software
(\textit{http://www.r-project.org}) (Fig. \ref{fig_locus}c). For computational reasons the genome is divided into several segments of equal size
and a separate model is trained on each segment (Supplementary Fig. \ref{Sfig_crossvalid}). The final energy of each word is the average over all models.
We restore the dynamic range of fitted energies by rescaling their variance to the variance of the Percus energies on which they were trained,
separately for each chromosome.
Finally, we predict nucleosome probabilities and occupancies from fitted energies using a standard recursive algorithm
(Fig. \ref{fig_locus}d; Supplementary Methods). \cite{segal:2006,morozov:2009}
Our predictions and software are available on the Nucleosome Explorer website, \textit{http://nucleosome.rutgers.edu}.

\section*{Acknowledgements}
This research was supported by National Institutes of Health (HG 004708 to A.V.M. and GM 30186 to K.S.).
A.V.M. acknowledges support from Alfred P. Sloan Research Fellowship.
We are grateful to Jerry K. Percus for helpful discussions.

\newpage
\section*{Figures}

\renewcommand{\topfraction}{.85}
\renewcommand{\bottomfraction}{.7}
\renewcommand{\textfraction}{.15}
\renewcommand{\floatpagefraction}{.66}
\renewcommand{\dbltopfraction}{.66}
\renewcommand{\dblfloatpagefraction}{.66}
\setcounter{topnumber}{9}
\setcounter{bottomnumber}{9}
\setcounter{totalnumber}{20}
\setcounter{dbltopnumber}{9}

\captionnamefont{\bfseries}
\captiontitlefont{\small\sffamily}
\renewcommand{\figurename}{Figure}
\captiondelim{. } 

\clearpage

\setcounter{figure}{0}
\renewcommand{\figurename}{Supplementary Figure}

%
%
%
%
%
%
%
%
%



\section*{Tables}

\begin{center}
\setlength{\belowcaptionskip}{6pt}   
\begin{table}[h]\small
\caption{ \textbf{Position-dependent distributions of sequence
motifs do not yield improved nucleosome occupancy predictions.
Shown are correlation coefficients between two \textit{in vitro}
nucleosome occupancy profiles \cite{zhang:2009,kaplan:2009} and
four $N=2$ models, and between the position-independent model and
three models with position-dependent energies.} }
\begin{tabular}{|c|c|c|c|c|} \hline
 & Position-independent & Three-region & Periodic & Spatially resolved \\ \hline \hline
Zhang \textit{et al.} & 0.60 & 0.62 & 0.59 & 0.61 \\ \hline
Kaplan \textit{et al.} & 0.75 & 0.75 & 0.74 & 0.75 \\ \hline
Position-independent & 1.00 & 0.99 & 1.00 & 0.99 \\ \hline 
\end{tabular}
\label{T1}
\end{table}
\end{center}
\clearpage

\setcounter{table}{0}
\renewcommand{\tablename}{Supplementary Table}

%



\newpage

\setcounter{section}{0}
\setcounter{equation}{0}
\renewcommand{\theequation}{S.\arabic{equation}}
\renewcommand{\thesection}{S.\arabic{section}}

{\center{\textbf{\LARGE Supplementary Material}}}

%
%

\section{Supplementary Methods}

\subsection{Preprocessing high-throughput sequencing data}

\noindent \textbf{Mapping sequence read profiles.}

We start from a collection of 25 bp-long Solexa sequence reads
uniquely mapped onto the yeast genome with no more than two
mismatches. \cite{zhang:2009} Each read is mapped onto either the
forward (5') or the reverse (3') strand. For sequence reads mapped
onto the forward (5') strand, we interpret the first base of a
read as the start position of a nucleosome with the canonical
length of 147 bp. For sequence reads mapped onto the reverse (3')
strand, we interpret the last base of the read as the end position
of a 147 bp nucleosome. Thus we create a ``sequence read
profile'', a table which shows the number of nucleosomes starting
at each genomic bp. This table is used to create a ``read coverage
profile'' which shows how many nucleosomes cover each genomic bp.

\noindent \textbf{Filtering sequence read profiles.}

We observe that there are large gaps in our read profiles,
possibly due to repetitive regions in the genome to which reads
cannot be mapped uniquely, or to sequencing artifacts. We
considered any stretch of $\ge 1000$ bp without mapped reads to be
anomalous and excluded such regions from further analysis. We also
find regions where the read coverage was uncharacteristically
high. For instance, our \textit{in vitro} nucleosome measurement
for chromosome 12 has an average nucleosome coverage of $\sim 80$
reads, but there is a small region near bp 460000 covered with
5000 reads. We exclude such regions according to the following
procedure: For each chromosome, we find the average number of
reads per bp. Next, for each bp we calculate the running average
number of reads in a window extending 75 bp in each direction. If
this running average is more than three times the mean, we flag
the region which extends out from the identified point in both
directions until the running average equals the mean, and we
remove this region from consideration. We then create a filter
which marks the union of all excluded regions. Finally, each
excluded region is extended 146 bp upstream so that there is no
contribution to the nucleosome energy from filtered regions.

\noindent \textbf{Normalizing sequence read profiles.}

Next we use the sequence read profile to create nucleosome
probability and occupancy profiles.
First, we set sequence read counts to zero inside all filtered
regions. Second, we use a Gaussian smoothing algorithm that
replaces the number of sequence reads at a given bp with a normal
distribution centered at that bp. The Gaussian is chosen to have
$\sigma=2$ or $20$ depending on subsequent modeling, and the area
under the curve is equal to the number of sequence reads at that
bp. The smoothed sequence read profile is then constructed as a
superimposition of all such Gaussians.

The smoothing procedure reflects a lack of bp precision in MNase
digestion assays, which results in the uncertainty of the
interpretation of sequence read coordinates as nucleosome start or
end positions. In addition, because neighboring nucleosomes are
expected to have similar binding affinities, collecting more
sequence read data is assumed to result in a read profile that we
approximate with the superposition of normal distributions
centered on available reads.

We extend the smoothed read profile into a smoothed read coverage
profile as described above, find the highest point $N_{max}$ in
the smoothed coverage profile and multiply the height of each
point in the smoothed coverage profile and the smoothed read
profile by $1/N_{max}$ so that the maximum coverage is one. Each
point in the smoothed sequence read profile may now be interpreted
as the probability for a nucleosome to start at a given position,
and the coverage may be interpreted as the probability for any
nucleosome to occupy a given position. We refer to the scaled
results as nucleosome probability and occupancy profiles,
respectively.

\subsection{Energetics of DNA-binding one-dimensional particles of finite size}

Consider particles of size $a$ bp distributed along a DNA segment
of length $L$ bp. The particles can interact with DNA in a
position-dependent manner and are also subject to steric exclusion
(adjacent particles cannot overlap). A grand-canonical partition
function for this system of DNA-bound particles is given by:

\begin{equation} \label{Peq:1}
Z = \sum_{conf} e^{-[E(conf) - \mu N(conf)]},
\end{equation}
where $conf$ denotes an arbitrary configuration of DNA-bound
non-overlapping particles, $\mu$ is the chemical potential, and
$E(conf)$ and $N(conf)$ are the total DNA-binding energy and the
number of particles in the current configuration (for simplicity
we assume $k_B T = 1$, where $k_B$ is the Boltzmann constant and
$T$ is the room temperature).

One can compute $Z$ efficiently using a recursive relation
\cite{morozov:2009}:

\begin{eqnarray} \label{Peq:2}
&&Z^f_i = Z^f_{i-1} + Z^f_{i-a} e^{-(E_{i-a+1} - \mu)},  \quad i = a, \dots, L \\
&&Z^f_{a-1} = \dots = Z^f_{0} = 1 \nonumber
\end{eqnarray}
which computes a set of partial partition functions in the forward
direction. Likewise, partial partition functions can be computed
in the reverse direction:

\begin{eqnarray} \label{Peq:3}
&&Z^r_i = Z^r_{i+1} + Z^r_{i+a} e^{-(E_{i} - \mu)},  \quad i = L-a+1, \dots, 1 \\
&&Z^r_{L-a+2} = \dots = Z^r_{L+1} = 1 \nonumber
\end{eqnarray}

Note that $Z^f_L = Z^r_1 = Z$ by construction. Furthermore, the
probability of starting a particle at position $i$ is given by:

\begin{equation} \label{Peq:4}
P_i = \frac{Z^f_{i-1} e^{-(E_{i} - \mu)} Z^r_{i+a}}{Z}, \quad i =
1, \dots, L-a+1
\end{equation}
Intuitively, Eq. (\ref{Peq:4}) is a partition function for all
configurations in which a particle is bound at position $i$
(occupying positions $i$ through $i+a-1$), divided by the
partition function for all possible configurations. Using Eqs.
(\ref{Peq:2}), (\ref{Peq:3}) and (\ref{Peq:4}) we obtain:

\begin{equation} \label{Peq:5}
\begin{array}{lc}
Z^f_i - Z^f_{i-1} = {P_{i-a+1} Z}/{Z^r_{i+1}}, & i = a, \dots, L \\
Z^r_{i+1} - Z^r_{i} = -{P_i Z}/{Z^f_{i-1}}, & i = L-a+1, \dots, 1
\end{array}
\end{equation}

Note that both of these formulas can be extended to the $i = 1,
\dots, L$ range if we assume that $P_k = 0, ~~k \notin [1,L-a+1]$.
It is easy to show that $Z^f_i  Z^r_{i+1} - Z^f_{i-1} Z^r_i  = Z
(P_{i-a+1} - P_i)$. This expression has the form of a complete
differential and thus can be iterated as follows:

\begin{equation} \label{Peq:6}
Z^f_L  Z^r_{L+1} - Z^f_{i-1} Z^r_i  = Z \sum_{j=i}^{L} (P_{j-a+1}
- P_j),
\end{equation}
yielding

\begin{equation} \label{Peq:7}
Z^f_{i-1} Z^r_i  = Z (1 - \sum_{j = i-a+1}^{i-1} P_j), \quad i =
1, \dots, L
\end{equation}
Using Eqs. (\ref{Peq:3}),  (\ref{Peq:4}) and  (\ref{Peq:7}) we
get:
\begin{equation} \label{Peq:8}
Z^r_{i+1} = Z^r_{i} \left( 1 - \frac{P_i}{1 - \sum_{j =
i-a+1}^{i-1} P_j} \right).
\end{equation}

Introducing $O_i = \sum_{j = i-a+1}^{i} P_j$ - the probability
that position $i$ is covered by any particle regardless of its
starting position (also called the particle occupancy), we see
that:
\begin{equation} \label{Peq:9}
Z^r_{i+1} = Z^r_{i} \left( 1 - \frac{P_i}{1 - O_i + P_i} \right).
\end{equation}
Using Eq. (\ref{Peq:9}) recursively (until $Z^r_{L+1} = 1$ is
reached on the left-hand side), we obtain an explicit expression
for $Z^r_{i}$:
\begin{equation} \label{Peq:10}
Z^r_{i} = \prod_{j=i}^{L} (1 - \frac{P_j}{1 - O_j + P_j})^{-1},
\quad  i = 1, \dots, L
\end{equation}
Likewise, using Eqs. (\ref{Peq:2}),  (\ref{Peq:4}) and
(\ref{Peq:7}) together with $Z^f_0 = 1$ we get:
\begin{equation} \label{Peq:11}
Z^f_{i} = \prod_{j=1}^{i} (1 - \frac{P_{j-a+1}}{1 - O_j +
P_{j-a+1}})^{-1}, \quad  i = 1, \dots, L
\end{equation}

Eqs.  (\ref{Peq:10}) and  (\ref{Peq:11}) are explicit expressions
for forward and reverse partial partition functions in terms of
particle probabilities and occupancies. Note that $Z^r_{1} =
Z^f_{L} = Z$ still holds, with Eqs. (\ref{Peq:10}) and
(\ref{Peq:11}) providing alternative expressions for the partition
function in this limit. Inserting  Eqs. (\ref{Peq:10}) and
(\ref{Peq:11}) into Eq. (\ref{Peq:4}) and using Eq. (\ref{Peq:7})
to express $Z^f_{i-1}$ in terms of $Z^r_{i}$ leads to the desired
expression for the DNA-binding energy of the particle at position
$i$:
\begin{equation} \label{Peq:12}
E_i - \mu = \log \frac{1 - O_i + P_i}{P_i} + \sum_{j=i}^{i+a-1}
\log \frac{1 - O_j}{1 - O_j + P_j}, \quad i = 1, \dots, L-a+1
\end{equation}
Alternatively, we can use  Eq. (\ref{Peq:7}) to express
$Z^r_{i+a}$ in terms of $Z^f_{i+a-1}$, leading to an equivalent
expression for the DNA-binding energy:
\begin{equation} \label{Peq:13}
E_i - \mu = \log \frac{1 - O_{i+a-1} + P_i}{P_i} +
\sum_{j=i-a+1}^{i} \log \frac{1 - O_{j+a-1}}{1 - O_{j+a-1} + P_j},
\quad i = 1, \dots, L-a+1
\end{equation}

\subsection{Hierarchical models of nucleosome energetics}

We have created hierarchical models of nucleosome energetics which
assign non-zero energies to nucleotide motifs of length $N$ only
if the nucleosome energies cannot be explained using nucleotide
motifs of lengths $1 \dots N-1$. This is implemented using
constraints on word energies:

\begin{equation} \label{Feq:1}
\sum_{\alpha_i} \epsilon_{\alpha_1 \dots \alpha_N} = 0, ~~\forall
i = 1 \dots N
\end{equation}
Here $\epsilon_{\alpha_1 \dots \alpha_N}$ is the energy of the
word of length $N$ with nucleotides $\alpha_1 \dots \alpha_N$ at
positions $1 \dots N$.

With these constraints and the $\{A,C,G,T\}$ alphabet there are
$3^N$ independent parameters describing energetics of words of
length $N$. For example, for $N = 1$ we can choose $\{ \epsilon_A,
\epsilon_G, \epsilon_T \}$ to be independent, while $\epsilon_C$
is fixed by the constraint: $\epsilon_C = - (\epsilon_A +
\epsilon_G + \epsilon_T)$. For $N=2$ there are 9 independent
parameters: $\{ \epsilon_{AA},  \epsilon_{AG}, \epsilon_{AT},
\epsilon_{GA}, \epsilon_{GG}, \epsilon_{GT}, \epsilon_{TA},
\epsilon_{TG}, \epsilon_{TT} \}$, while the other 7 dinucleotide
energies can be expressed through these using Eq. (\ref{Feq:1}).
The remaining 7 degrees of freedom are described by the lower
order terms: 6 $\epsilon_{\alpha}$'s (3 for each position in the
dinucleotide) and the total offset $\epsilon^0$.

In general, $D^N$ degrees of freedom associated with words of
length $N$ drawn from an alphabet of size $D$ can be described
using constrained energies:
\begin{equation} \label{Feq:2}
D^N = (D-1)^N + \binom{N}{1} (D-1)^{N-1} + \dots + \binom{N}{N}
(D-1)^{0},
\end{equation}
where each term describes the total number of constrained energies
of order $(N, \dots ,0)$, computed as a product of the number of
constrained energies at each possible position within the longer
word, and the number of such positions. Note that the zeroth order
term is simply the total offset $\epsilon^0$. Furthermore, shorter
words comprised of non-consecutive nucleotides are included in the
expansion. If we set the energies of all non-consecutive words to
zero, the total energy of a word of length $N$ can be written as:
\begin{equation} \label{Feq:3}
\epsilon'_{\alpha_1 \dots \alpha_N} = \sum_{n=1}^{N} \sum_{j=1}^{N
- n + 1} \epsilon_{\alpha_j \dots \alpha_{j+n-1}} + \epsilon^0
\end{equation}

Note that here and in Section \ref{ssmodels} below we set $\mu =
0$ for simplicity. Although a set of constrained energies of order
$0, \dots ,N$ on the right-hand side of Eq. (\ref{Feq:3}) has
fewer degrees of freedom than a set of unconstrained energies of
order $N$, it provides the most complete description involving
consecutive nucleotide motifs, and forms a basis of nucleosome
models that have been further simplified by equating energies of
motifs that occur at different positions within the nucleosomal
site. Furthermore, since dinucleotides are too short to contain
partial non-consecutive motifs, Eq. (\ref{Feq:3}) entails no loss
of degrees of freedom for $N=2$.

\subsection{Sequence-specific models of nucleosome energetics} \label{ssmodels}

Eq. (\ref{Peq:12}) can be used to convert nucleosome probabilities
and occupancies obtained from high-throughput sequencing data into
histone-DNA interaction energies for each position $i$ along the
DNA, under the assumption that steric exclusion and specific
interactions with DNA are the only factors that affect nucleosome
positions \textit{in vitro}. In order to understand which DNA
sequence features explain the observed energy profile, we carried
out linear fits of genome-wide Percus energies (Eq.
(\ref{Peq:12})) to four sequence-specific models. Some models were
designed to focus on the $\sim \! 10-11$ bp periodic distributions
of sequence motifs, while others capture nucleosome-wide sequence
signals such as motif enrichment and depletion in
nucleosome-covered sequences.

\noindent \emph{Spatially resolved model.}

In terms of unconstrained energies, the spatially resolved model
is defined as:
\begin{equation} \label{Meq:1}
E(S) = \sum_{i=I_1}^{I_2-1} \epsilon'_{\alpha_i \alpha_{i+1}},
\end{equation}
where $E(S)$ is the nucleosome formation energy of a 147 bp-long
sequence $S$, $\epsilon_{\alpha_{i} \alpha_{i+1}}$ is the energy
of the dinucleotide with bases $\alpha_i$ and $\alpha_{i+1}$ at
positions $i$ and $i+1$ respectively, and the sum runs from $I_1
\ge 1$ to $I_2 \le 147$ in the nucleosomal site. To minimize edge
effects, we typically exclude 3 bps from each end of the
nucleosome, setting $I_1 = 4$ and $I_2 = 144$.

Eq. (\ref{Meq:1}) can be rewritten as:
\begin{equation} \label{Meq:2}
E(S) = \sum_{i=I_1}^{I_2-1} \left( \epsilon_{\alpha_i
\alpha_{i+1}} + \bar{b}_{\alpha_i} + b_{\alpha_{i+1}} \right) +
\epsilon^0,
\end{equation}
where
\begin{eqnarray*}
\epsilon^0 &=&  \frac{1}{D^2} \sum_{i=I_1}^{I_2-1} \sum_{\alpha,\beta = 1}^{D} \epsilon'_{\alpha \beta} \equiv \sum_{i=I_1}^{I_2-1} \epsilon^0_{i,i+1} , \\
\bar{b}_{\alpha} &=& \frac{1}{D} \sum_{\beta = 1}^{D} \left( \epsilon'_{\alpha \beta} - \epsilon^0_{i,i+1} \right), \\
b_{\beta} &=&  \frac{1}{D} \sum_{\alpha = 1}^{D} \left(
\epsilon'_{\alpha \beta} - \epsilon^0_{i,i+1} \right).
\end{eqnarray*}

Note that $\sum_{\alpha = 1}^{D} \epsilon_{\alpha \beta} =
\sum_{\beta = 1}^{D} \epsilon_{\alpha \beta} = 0$ by construction.
Eq. (\ref{Meq:2}) is equivalent to the expansion in terms of
constrained energies which is consistent with Eq. (\ref{Feq:3}):
\begin{equation} \label{Meq:3}
E(S) = \sum_{i=I_1}^{I_2-1} \epsilon_{\alpha_i \alpha_{i+1}} +
\sum_{i=I_1}^{I_2} \epsilon_{\alpha_i} + \epsilon^0,
\end{equation}
where $\epsilon_{\alpha_{I_1}} = \bar{b}_{\alpha_{I_1}},
~\epsilon_{\alpha_{I_1+1}} = \bar{b}_{\alpha_{I_1+1}} +
b_{\alpha_{I_1+1}}, \dots , ~\epsilon_{\alpha_{I_2}} =
b_{\alpha_{I_2}}$. Thus an unconstrained description of nucleosome
energetics can be uniquely decomposed into a constrained
description. However, the opposite is not true: for any $p$ and
$q$ such that $p + q = 1$
\begin{equation} \label{Meq:3b}
\begin{array}{lcl}
\epsilon'_{\alpha_{I_1} \alpha_{I_1+1}} &=& \epsilon_{\alpha_{I_1} \alpha_{I_1+1}} + \epsilon_{\alpha_{I_1}} + q \epsilon_{\alpha_{I_1+1}},  \\
\epsilon'_{\alpha_i \alpha_{i+1}} &=& \epsilon_{\alpha_i
\alpha_{i+1}} + p \epsilon_{\alpha_i} + q \epsilon_{\alpha_{i+1}},
~~I_1 < i < I_2-1 \\ \nonumber \epsilon'_{\alpha_{I_2 - 1}
\alpha_{I_2}} &=& \epsilon_{\alpha_{I_2 - 1} \alpha_{I_2}} + p
\epsilon_{\alpha_{I_2 - 1}} + \epsilon_{\alpha_{I_2}} \nonumber
\end{array}
\end{equation}
are equally valid reconstructions that leave $E(S)$ unchanged. In
this paper we use $p=1, q=0$ to compute unconstrained dinucleotide
energies from constrained ones.

\noindent \emph{Position-independent model.}

This model assigns the same energy to a given word within the
nucleosome, regardless of its position in the site. Thus the
position-independent model of order $N$ is given by:
\begin{equation} \label{Meq:4}
E(S) = \sum_{n = 1}^{N} \sum_{\{\alpha_{1} \dots
\alpha_{n}\}}^{3^n} n_{\alpha_{1} \dots \alpha_{n}}
\epsilon_{\alpha_{1} \dots \alpha_{n}} + \epsilon^0,
\end{equation}
where the outer sum is over word lengths, the inner sum is over
all words of length $n$ corresponding to constrained energies,
$n_{\alpha_{1} \dots \alpha_{n}}$ is the number of words with the
nucleotides $\alpha_1 \dots \alpha_n$ at positions $1 \dots n$,
and  $\epsilon_{\alpha_{1} \dots \alpha_{n}}$ are word energies
constrained by Eq. (\ref{Feq:1}). As in the spatially resolved
model, the words are counted from bp $I_1 = 4$ to bp $I_2 = 144$,
excluding 3 bp from each end of the site. The words are not
allowed to extend outside this region. Note that both in this
model and in the two partially position-dependent models described
below there is no one-to-one correspondence between constrained
models utilizing words of order $1 \dots N$ and their
unconstrained counterparts utilizing words of order $N$ - the
former require fewer fitting parameters.

\noindent \emph{Three-region model.}

This model refines the position-independent model by dividing the
141 bp nucleosome site into 3 regions of equal length. Word
energies are fitted separately inside each region. The total
energy of sequence $S$ is then given by:
\begin{equation} \label{Meq:5}
E(S) = \sum_{r = 1}^{3} \sum_{n = 1}^{N} \sum_{\{\alpha_{1} \dots
\alpha_{n}\}}^{3^n} n^{r}_{\alpha_{1} \dots \alpha_{n}}
\epsilon^{r}_{\alpha_{1} \dots \alpha_{n}} + \epsilon^0,
\end{equation}
where $r$ refers to a particular 47 bp region.

\noindent \emph{Periodic model.}

This model enforces DNA helical twist periodicity by equating the
energies of words separated by a multiple of 10 bp. To reduce the
number of fitting parameters, we also grouped energies of words at
positions $1 \dots 10$ into 5 distinct bins. Thus an AGT motif
starting at position 1 within the nucleosome site would have the
same energy as the AGT motif starting at positions $11, 21, 31
\dots$ as well as positions $2, 12, 22 \dots$, whereas the energy
of the same motif starting at positions 3 and 4 is grouped into a
different bin. The total energy is then computed as:

\begin{equation} \label{Meq:6}
E(S) = \sum_{b = 1}^{5} \sum_{n = 1}^{N} \sum_{\{\alpha_{1} \dots
\alpha_{n}\}}^{3^n} n^{b}_{\alpha_{1} \dots \alpha_{n}}
\epsilon^{r}_{\alpha_{1} \dots \alpha_{n}} + \epsilon^0,
\end{equation}
where $b$ is the bin index used to group motifs separated by the
helical twist as described above. As before, all words overlapping
with the 3 bp edge regions are excluded from the counts.

\newpage
\section*{Supplementary Figures}

\renewcommand{\topfraction}{.85}
\renewcommand{\bottomfraction}{.7}
\renewcommand{\textfraction}{.15}
\renewcommand{\floatpagefraction}{.66}
\renewcommand{\dbltopfraction}{.66}
\renewcommand{\dblfloatpagefraction}{.66}
\setcounter{topnumber}{9} \setcounter{bottomnumber}{9}
\setcounter{totalnumber}{20} \setcounter{dbltopnumber}{9}

\captionnamefont{\bfseries} \captiontitlefont{\small\sffamily}
\renewcommand{\figurename}{Supplementary Figure}
\captiondelim{. } 


\begin{figure}[tbph]
\centering
\includegraphics[scale=0.92]{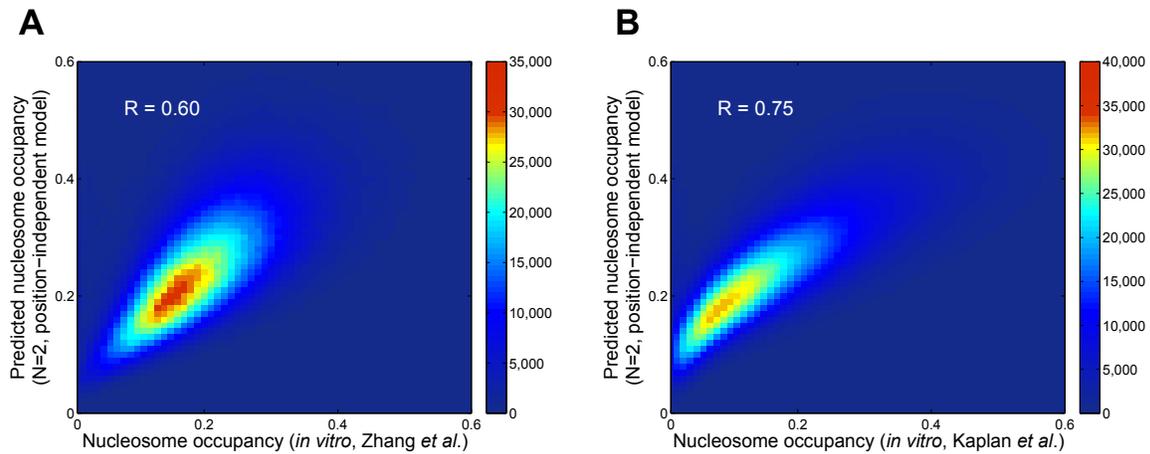}
\caption{ \label{Sfig_N2} \textbf{$N=2$ position-independent model
is sufficient to explain nucleosome occupancy in
\textit{S.cerevisiae}.} a) Density scatter plot for the nucleosome
occupancy at each genomic base pair (predicted with the $N=2$
position-independent model) vs. \textit{in vitro} occupancy
observed by Zhang \textit{et al.} \cite{zhang:2009} b) Same as (a)
except that \textit{in vitro} occupancy is from Kaplan \textit{et
al.} \cite{kaplan:2009} }
\end{figure}

\begin{figure}[tbph]
\centering
\includegraphics[scale=0.50]{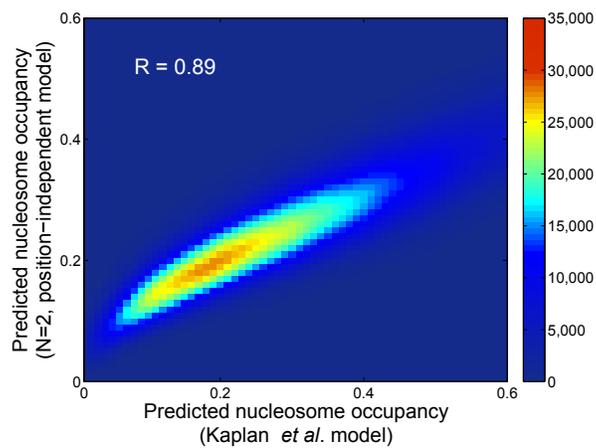}
\caption{ \label{Sfig_Kaplan_model} \textbf{Similar predictive
power of the $N=2$ position-independent model and a bioinformatics
model based on periodic dinucleotide distributions and frequencies
of 5 bp-long words. \cite{kaplan:2009}} Density scatter plot for
the nucleosome occupancy at each genomic base pair (predicted with
the $N=2$ position-independent model) vs. nucleosome occupancy
predicted by Kaplan \textit{et al.} \cite{kaplan:2009} }
\end{figure}

\begin{figure}[tbph]
\centering
\includegraphics[angle=-90,scale=0.85]{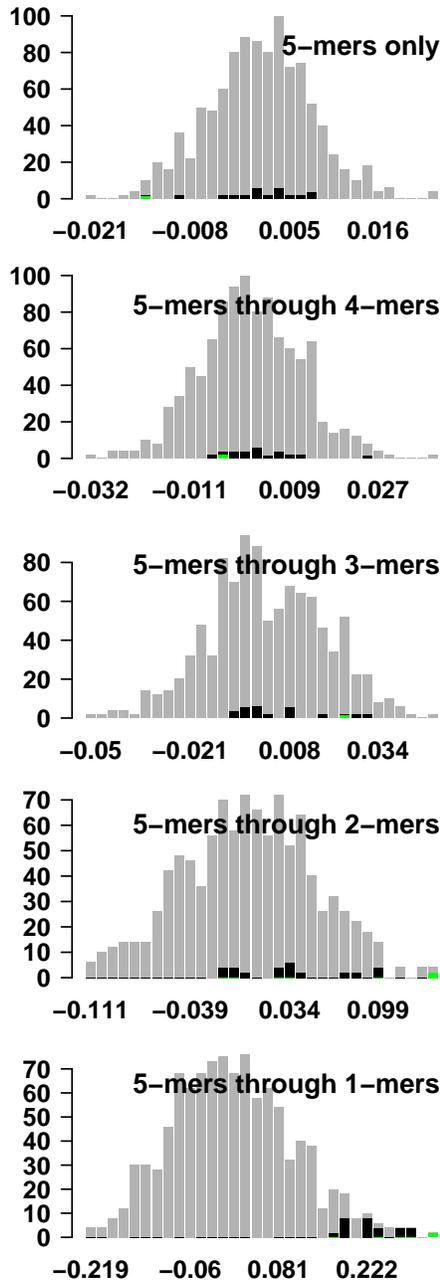}
\caption{ \label{Sfig_polyA} \textbf{Minor role of the
higher-order contributions to the energies of 5 bp-long words.}
$N=5$ position-independent model was trained on nucleosomes
reconstituted \textit{in vitro} on the yeast genome,
\cite{zhang:2009} yielding energies of all motifs of 1 through 5
bp in length. Energies of 5 bp-long words were then computed by
summing contributions from a subset of shorter motifs: $E(S) =
\sum_{n=L}^5 \sum_{\{\alpha_1 \dots \alpha_n\}}^{4^n} n_{\alpha_1
\dots \alpha_n} \epsilon_{\alpha_1 \dots \alpha_n}$, where
$n_{\alpha_1 \dots \alpha_n}$ is the number of times a given word
was found in the 5 bp-long sequence $S$ and $\epsilon_{\alpha_1
\dots \alpha_n}$ is the fitted energy of that word. $L = 5 \dots
1$ is the length of the shortest motif included into $E(S)$. Grey:
all 5 bp-long words, black: A:T-containing words, green: the
poly(dA:dT) tract (AAAAA). }
\end{figure}

\begin{figure}[tbph]
\centering
\includegraphics[scale=0.90]{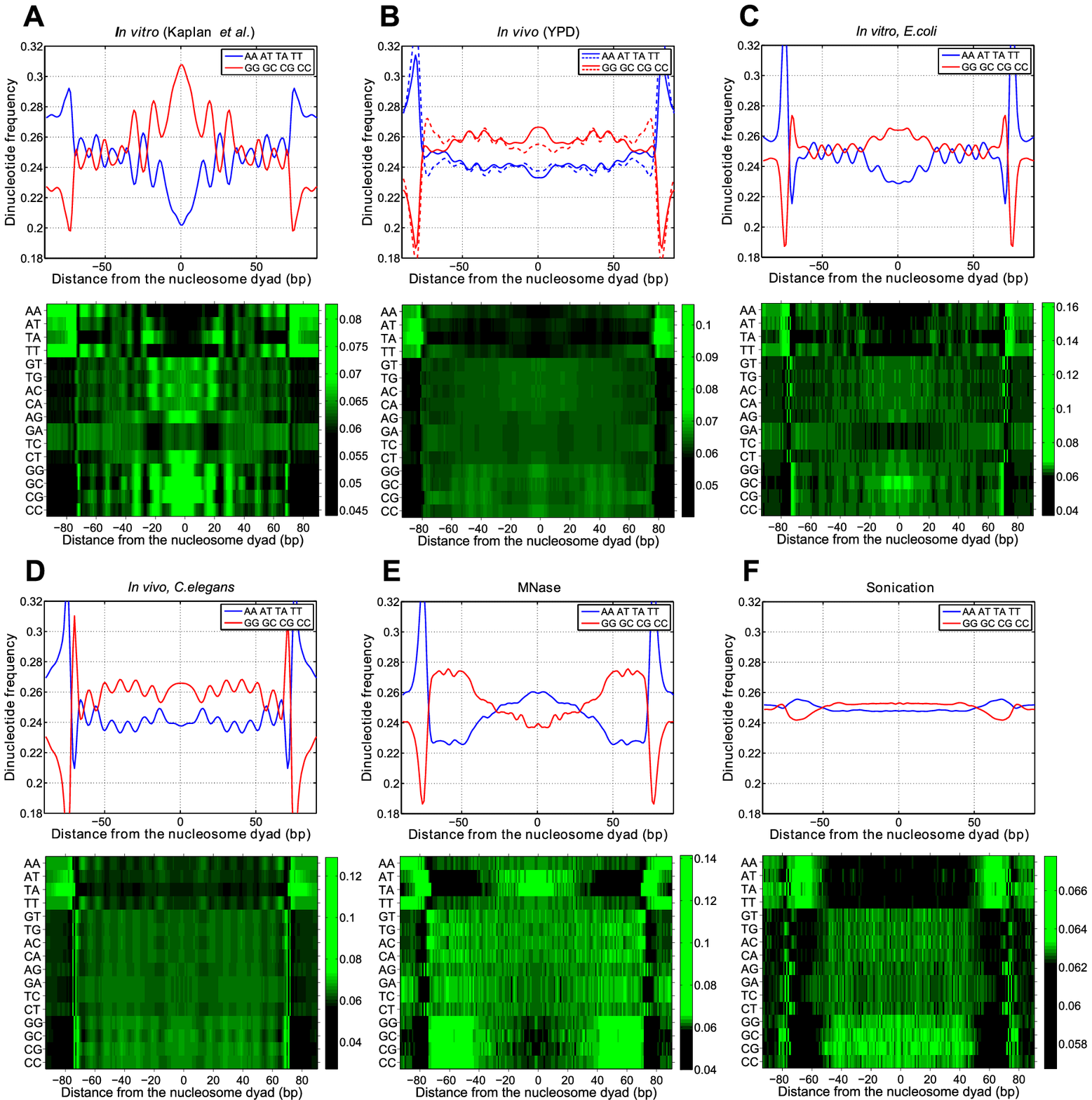}
\end{figure}


\begin{figure}[tbph]
\centering \caption{ \label{Sfig_green_blue_red}
\textbf{Dinucleotide distributions in nucleosome and linker
sequences.} \emph{Upper panel}: average relative frequencies of WW
(AA, TT, AT and TA) and SS (CC, GG, CG and GC) dinucleotides at
each position within the nucleosome are plotted with respect to
the nucleosome dyad. The relative frequency of each dinucleotide
is defined as its frequency at a given position divided by
genome-wide frequency.
All frequencies are smoothed using a 3 bp moving average.
\emph{Lower panel}: heat map of relative frequencies for each
dinucleotide, plotted with respect to the nucleosome dyad.
a) Nucleosomes assembled \textit{in vitro} on the yeast genome
(defined by more than five sequence reads), from Kaplan \textit{et
al.} \cite{kaplan:2009} b) \textit{In vivo} nucleosomes (defined
by more than five sequence reads) from yeast cells grown in YPD
medium. \cite{kaplan:2009} Upper panel: dashed lines -
cross-linked nucleosomes, solid lines - no cross-linking. Lower
panel: dinucleotide counts based on a combination of all YPD
replicates. c) Nucleosomes assembled \textit{in vitro} on the
\textit{E.coli} genome (defined by more than one sequence read).
\cite{zhang:2009} d) \textit{In vivo} nucleosomes (defined by more
than three sequence reads) from \textit{C.elegans}.
\cite{valouev:2008} e) Same as (a)-(d) except the dinucleotide
frequencies are from mononucleosome-size DNA sequences (defined by
more than five sequence reads) from yeast genomic DNA digested by
MNase in the absence of nucleosomes. f) Same as (e) except
mononucleosome-size DNA sequences (defined by more than one
sequence read) were obtained by sonication. }
\end{figure}

\begin{figure}[tbph]
\centering
\includegraphics[scale=0.85]{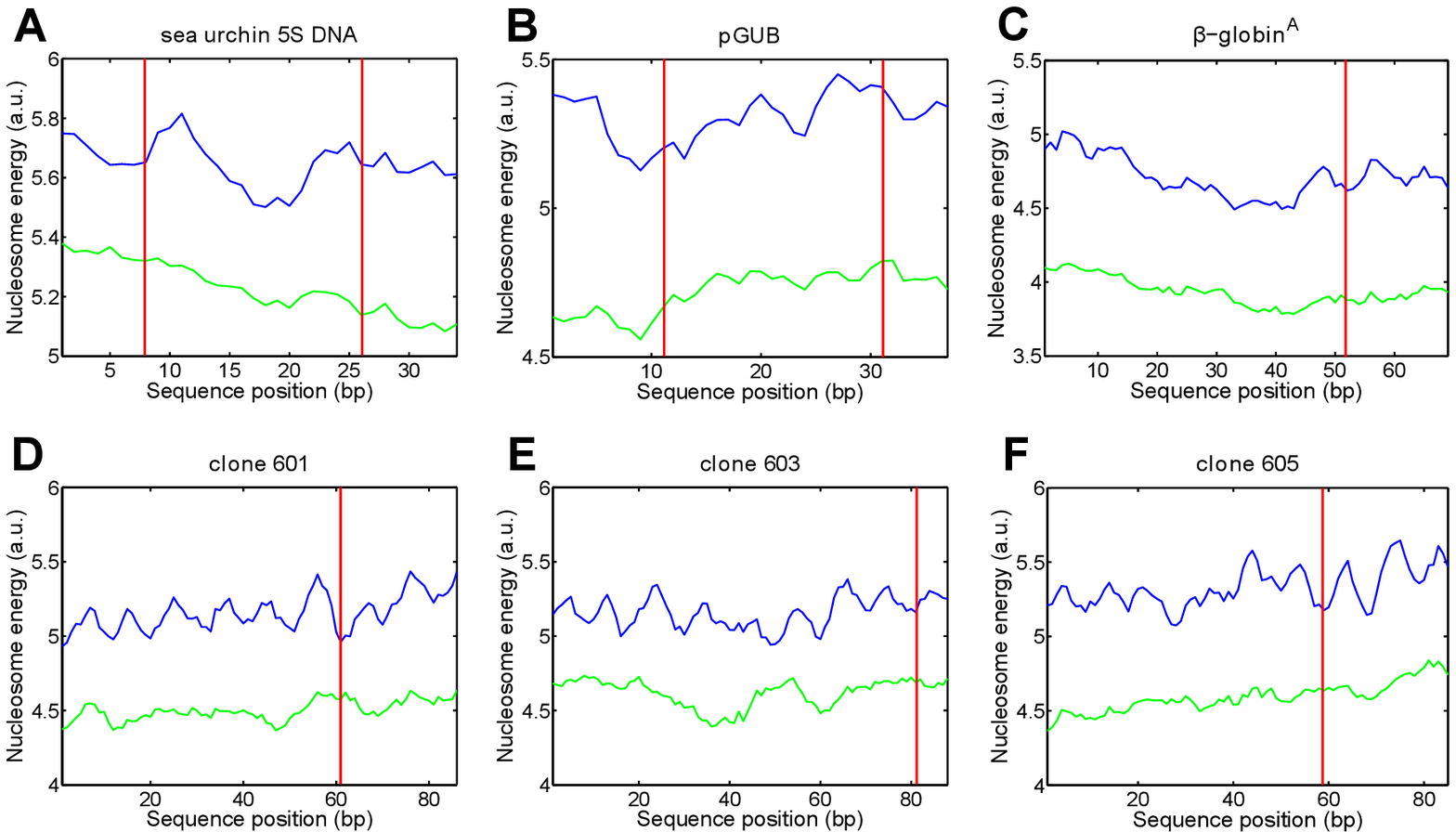}
\caption{ \label{Sfig_6nucs} \textbf{Prediction of six nucleosome
positions mapped \textit{in vitro} at high resolution.} Shown are
nucleosome formation energies computed using the $N=2$
position-independent model (green curves) and the spatially
resolved model (blue curves). Vertical lines: known nucleosome
starting positions, also listed in parentheses below. (a) The 180
bp sequence from the sea urchin 5S rRNA gene (bps 8,26).
\cite{flaus:1996} (b) The 183 bp sequence from the pGUB plasmid
(bps 11,31). \cite{kassabov:2002} (c) The 215 bp fragment from the
sequence of the chicken $\beta-\mathrm{globin}^{\mathrm{A}}$ gene
(bp 52). \cite{davey:2004} (d,e,f) Synthetic high-affinity
sequences \cite{lowary:1998} 601 (bp 61), 603 (bp 81) and 605 (bp
59). \cite{morozov:2009} }
\end{figure}

\begin{figure}[tbph]
\centering
\includegraphics[scale=0.80]{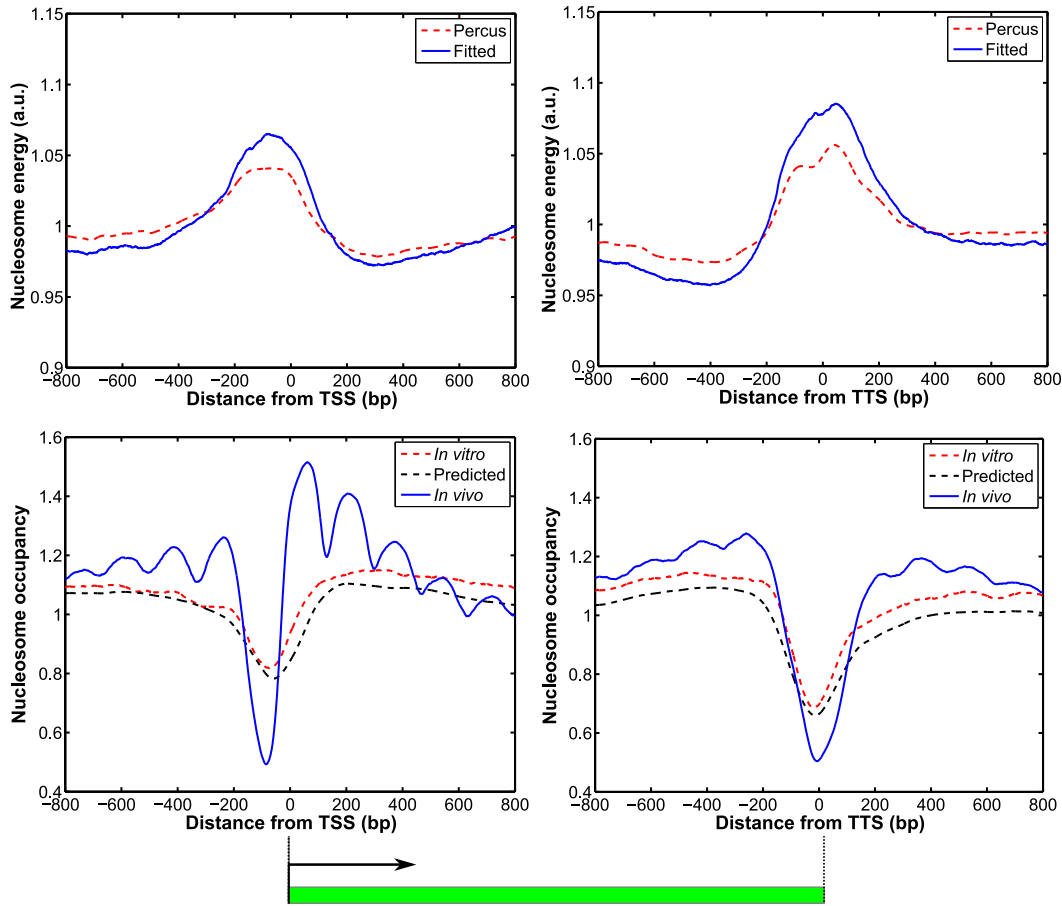}
\caption{ \label{Sfig_TSS_TTS} \textbf{Nucleosome energies and
occupancies in the vicinity of transcription start and termination
sites.} a) Percus energy (red) and the sequence-specific energy
predicted using the $N=2$ position-independent model (blue). The
energies were inferred from nucleosomes positioned \textit{in
vitro} on the yeast genome,  \cite{zhang:2009} averaged over all
genes for which transcript coordinates were available,
\cite{nagalakshmi:2008} and plotted with respect to the
transcription start and termination sites (TSS and TTS,
respectively). All energies were divided by a genome-wide average.
b) \textit{In vitro} nucleosome occupancy (red), \cite{zhang:2009}
\textit{in vivo} nucleosome occupancy in YPD medium without
cross-linking (blue), \cite{kaplan:2009} and occupancy predicted
using the $N=2$ position-independent model (black). All
occupancies were divided by the genome-wide average and plotted as
described in (a). }
\end{figure}

\begin{figure}[tbph]
\centering
\includegraphics[scale=0.65]{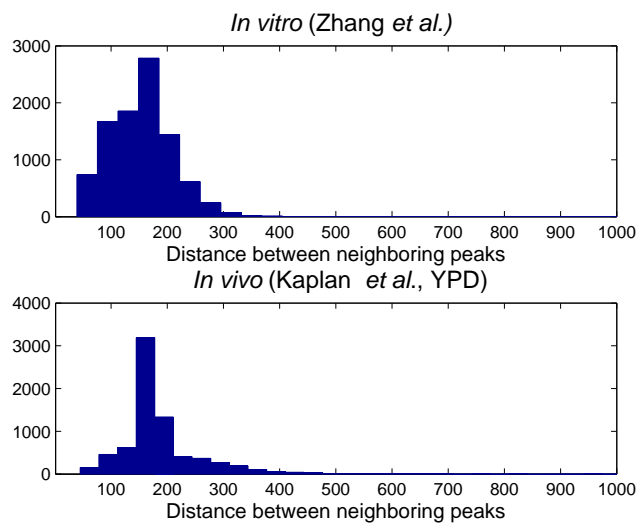}
\caption{ \label{Sfig_disthist} \textbf{Histogram of distances
between neighboring peaks from \textit{in vitro} and \textit{in
vivo} nucleosome sequence read profiles in \textit{S.cerevisiae}.}
Mapped sequence reads were smoothed with a $\sigma=20$ Gaussian.
Neighboring peaks are defined by local maxima in the  sequence
read profile. }
\end{figure}

\begin{figure}[tbph]
\centering
\includegraphics[scale=0.50]{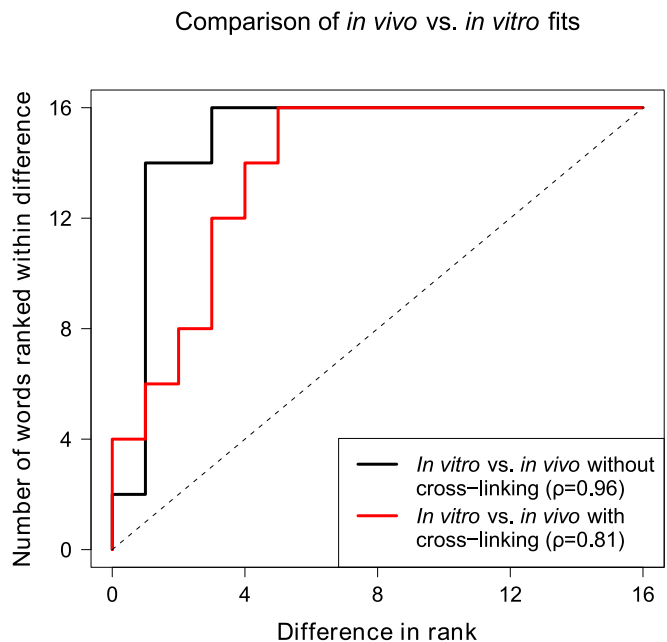}
\caption{ \label{Sfig_invivo} \textbf{Comparison of $N=2$
position-independent models trained on \textit{in vitro} and
\textit{in vivo} \textit{S.cerevisiae} nucleosomes.}
Rank-order plots of energies of 2 bp words: the energy of each
word is ranked using a position-independent model of order $N=2$
trained on either \textit{in vivo} (with and without
cross-linking) or \textit{in vitro} nucleosome positioning data.
Each curve shows the number of words whose ranks are separated in
the \textit{in vivo} vs. \textit{in vitro} fits by a given
distance or less. }
\end{figure}

\begin{figure}[tbph]
\centering
\includegraphics[scale=0.55,angle=-90]{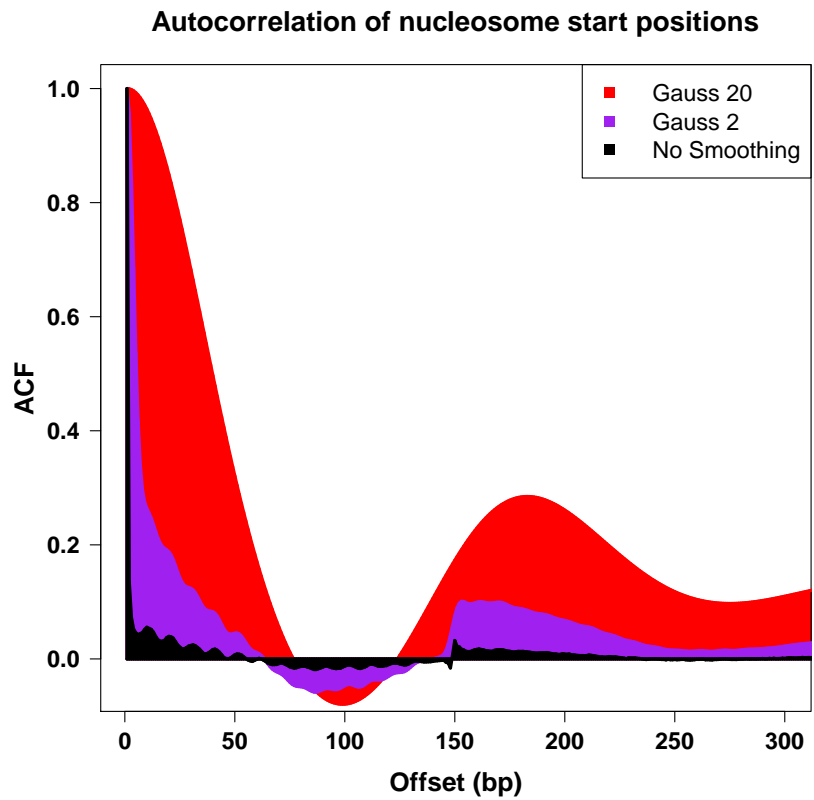}
\caption{ \label{Sfig_acf} \textbf{Autocorrelation functions of
nucleosome starting positions.} Nucleosomes were assembled
\textit{in vitro} on the yeast genome. \cite{zhang:2009} Black:
original starting positions, violet: starting positions smoothed
with a $\sigma=2$ Gaussian, red: starting positions smoothed with
a $\sigma=20$ Gaussian (see Supplementary Methods). }
\end{figure}

\begin{figure}[tbph]
\centering
\includegraphics[scale=0.90]{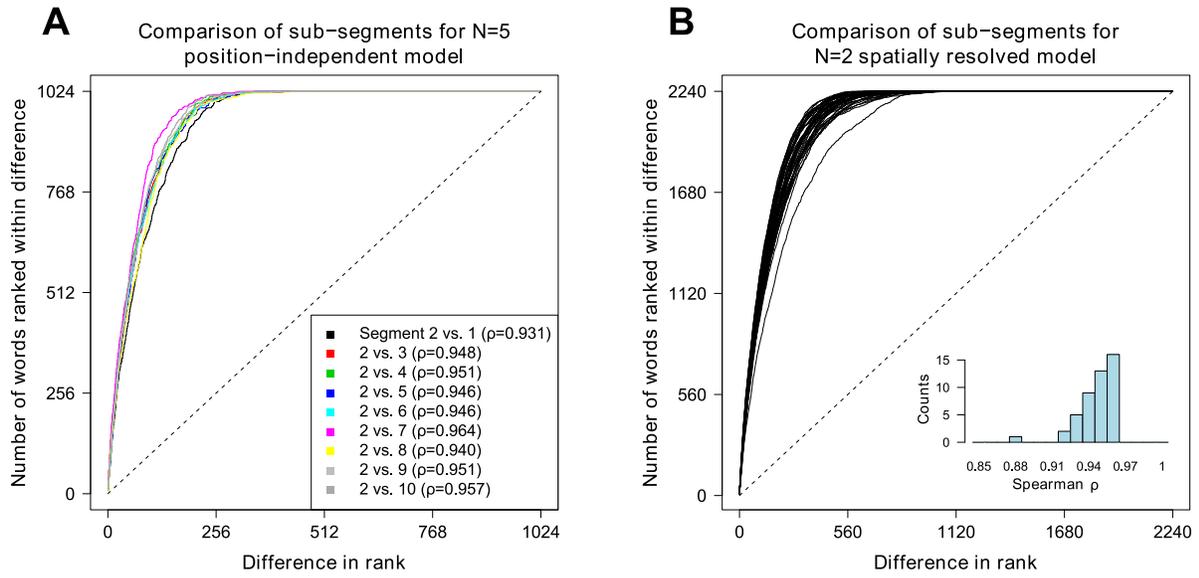}
\caption{ \label{Sfig_crossvalid} \textbf{Cross-validation of the
$N=5$ position-independent and $N=2$ spatially resolved models in
\textit{S.cerevisiae}.} a) Rank-order plots of energies of 5 bp
words: yeast genome is divided into 4 segments of equal size and
the energy of each word is ranked using $N=5$ position-independent
models independently trained on each segment. Each curve shows the
number of words whose ranks are separated by a given distance or
less. Energies of 5 bp-long words contain contributions from all
shorter motifs: $E(S) = \sum_{n=1}^5 \sum_{\{\alpha_1 \dots
\alpha_n\}}^{4^n} n_{\alpha_1 \dots \alpha_n} \epsilon_{\alpha_1
\dots \alpha_n}$, where $n_{\alpha_1 \dots \alpha_n}$ is the
number of times a given word was found within the 5 bp-long
sequence $S$ and $\epsilon_{\alpha_1 \dots \alpha_n}$ is the
fitted energy of that word. b) Rank-order plots of dinucleotide
energies at each position predicted with $N=2$ spatially resolved
models independently trained on 47 segments of equal size.
Dinucleotide energies at each position are computed using
$E_{\alpha_{i} \alpha_{i+1}} = \epsilon_{\alpha_i \alpha_{i+1}} +
\epsilon_{\alpha_i}, ~i=4 \dots 142$, $E_{\alpha_{143}
\alpha_{144}} = \epsilon_{\alpha_{143} \alpha_{144}} +
\epsilon_{\alpha_{143}} + \epsilon_{\alpha_{144}}$ (Supplementary
Methods) and ranked across all positions. The inset shows a
histogram of rank-order correlation coefficients between
dinucleotide energies trained on one of the segments, and all
other segments. }
\end{figure}
\clearpage


\section*{Supplementary Tables}

\renewcommand{\tablename}{Supplementary Table}

\begin{center}
\setlength{\belowcaptionskip}{6pt}   
\begin{table}[h]
\caption{ \textbf{Table of correlation coefficients between
predicted or observed occupancy profiles on the yeast genome.} All
observed profiles have been filtered for abnormally high- and
low-density regions as described in the Supplementary Methods,
with each correlation coefficient computed only for those
basepairs that have not been removed from either dataset
(predicted occupancies do not have filtered regions). The table is
available at
\href{http://nucleosome.rutgers.edu}{http://nucleosome.rutgers.edu}.
} \label{ST1}
\end{table}
\end{center}

\begin{center}
\setlength{\belowcaptionskip}{6pt}   
\begin{table}[h]
\caption{ \textbf{Table of dinucleotide energies predicted by
training $N=2$ position-independent models on several nucleosome
positioning maps and nucleosome-free control experiments.}
Energies for each model have been rescaled to the variance of 1
a.u. } \label{ST2}\footnotesize
\begin{tabular}{|r|r|r|r|r|r|r|r|r|r|r|}
\hline
Species/Control & \multicolumn{2}{|c|}{S. cerevisiae} & \multicolumn{2}{|c|}{E. Coli} & \multicolumn{2}{|c|}{C. Elegans} & \multicolumn{2}{|c|}{MNase } & \multicolumn{2}{|c|}{Sonicated}\\
\hline
rank & word & energy & word & energy & word & energy & word & energy & word & energy\\
\hline\hline
1 & TT & 1.76 & TT & 2.14 & TT & 1.57 & TT & 1.40 & AT & 1.30\\
\hline
2 & AA & 1.76 & AA & 2.14 & AA & 1.57 & AA & 1.40 & TA & 1.30\\
\hline
3 & TA & 1.10 & TA & 0.59 & CG & 1.36 & AT & 0.90 & TT & 1.07\\
\hline
4 & AT & 0.98 & CT & 0.26 & GC & 0.93 & GA & 0.62 & AA & 1.07\\
\hline
5 & CT & 0.27 & AG & 0.26 & TA & 0.71 & TC & 0.62 & CT & 0.31\\
\hline
6 & AG & 0.27 & AT & 0.25 & CC & 0.40 & AG & 0.32 & AG & 0.31\\
\hline
7 & TC & 0.19 & GG & 0.09 & GG & 0.40 & CT & 0.32 & GA & 0.16\\
\hline
8 & GA & 0.19 & CC & 0.09 & AT & 0.06 & TA & 0.32 & TC & 0.16\\
\hline
9 & AC & -0.50 & GC & -0.15 & AG & -0.69 & TG & 0.04 & TG & 0.05\\
\hline
10 & GT & -0.50 & GA & -0.36 & CT & -0.69 & CA & 0.04 & CA & 0.05\\
\hline
11 & CA & -0.55 & TC & -0.36 & GT & -0.73 & GT & -0.24 & AC & -0.09\\
\hline
12 & TG & -0.55 & TG & -0.84 & AC & -0.73 & AC & -0.24 & GT & -0.09\\
\hline
13 & GG & -0.81 & CA & -0.84 & GA & -0.80 & CC & -0.77 & GG & -0.86\\
\hline
14 & CC & -0.81 & CG & -1.04 & TC & -0.80 & GG & -0.77 & CC & -0.86\\
\hline
15 & GC & -1.40 & AC & -1.12 & TG & -1.28 & CG & -1.97 & GC & -1.79\\
\hline
16 & CG & -1.42 & GT & -1.12 & CA & -1.28 & GC & -1.99 & CG & -2.09\\
\hline
\end{tabular}

\end{table}
\end{center}

\newpage
\bibliographystyle{nar}
\bibliography{nucref}

\begin{thebibliography}{10}

\bibitem{vanholde:1989}
{van Holde}, K.~E. (1989)
Chromatin,
Springer, New York.

\bibitem{richmond:2003}
Richmond, T.~J. and Davey, C.~A. (2003)
{\em Nature} {\bf 423}, 145--150.

\bibitem{felsenfeld:2003}
Felsenfeld, G. and Groudine, M. (2003)
{\em Nature} {\bf 421}, 448--453.

\bibitem{jenuwein:2001}
Jenuwein, T. and Allis, C. (2001)
{\em Science} {\bf 293}, 1074--1080.

\bibitem{kaplan:2009}
Kaplan, N., Moore, I.~K., Fondufe-Mittendorf, Y., Gossett, A.~J., Tillo, D.,
  Field, Y., LeProust, E.~M., Hughes, T.~R., Lieb, J.~D., Widom, J., and Segal,
  E. (2009)
{\em Nature} {\bf 458}, 362--366.

\bibitem{sekinger:2005}
Sekinger, E.~A., Moqtaderi, Z., and Struhl, K. (2005)
{\em Mol. Cell} {\bf 18}, 735--748.

\bibitem{becker:2002}
Becker, P.~B. and H\"{o}rz, W. (2002)
{\em Annu. Rev. Biochem.} {\bf 71}, 247--273.

\bibitem{deckert:2001}
Deckert, J. and K., S. (2001)
{\em Mol. Cell. Biol.} {\bf 21}, 2726--2735.

\bibitem{adams:1995}
Adams, C.~C. and Workman, J.~L. (1995)
{\em Mol. Cell. Biol.} {\bf 15}, 1405--1421.

\bibitem{miller:2003}
Miller, J.~A. and Widom, J. (2003)
{\em Mol. Cell. Biol.} {\bf 23}, 1623--1632.

\bibitem{segal:2006}
Segal, E., Fondufe-Mittendorf, Y., Chen, L.~Y., Thastrom, A., Field, Y., Moore,
  I.~K., Wang, J. P.~Z., and Widom, J. (2006)
{\em Nature} {\bf 442}, 772--778.

\bibitem{kornberg:1988}
Kornberg, R.~D. and Stryer, L. (1988)
{\em Nucleic Acids Res.} {\bf 16}, 6677--6690.

\bibitem{fedor:1988}
Fedor, M.~J., Lue, N.~F., and Kornberg, R.~D. (1988)
{\em J. Mol. Biol.} {\bf 204}, 109--127.

\bibitem{mavrich:2008}
Mavrich, T.~N., Jiang, C.~Z., Ioshikhes, I.~P., Li, X.~Y., Venters, B.~J.,
  Zanton, S.~J., Tomsho, L.~P., Qi, J., Glaser, R.~L., Schuster, S.~C.,
  Gilmour, D.~S., Albert, I., and Pugh, B.~F. (2008)
{\em Nature} {\bf 453}, 358--362.

\bibitem{mavrich:2008b}
Mavrich, T.~N., Ioshikhes, I.~P., Venters, B.~J., Jiang, C., Tomsho, L.~P., Qi,
  J., Schuster, S.~C., Albert, I., and Pugh, B.~F. (2008)
{\em Genome Res.} {\bf 18}, 1073--1083.

\bibitem{widom:2001}
Widom, J. (2001)
{\em Q. Rev. Biophys.} {\bf 34}, 269--324.

\bibitem{ioshikhes:2006}
Ioshikhes, I.~P., Albert, I., Zanton, S.~J., and Pugh, B.~F. (2006)
{\em Nature Genet.} {\bf 38}, 1210--1215.

\bibitem{field:2008}
Field, Y., Kaplan, N., Fondufe-Mittendorf, Y., Moore, I.~K., Sharon, E.,
  Lubling, Y., Widom, J., and Segal, E. (2008)
{\em PLoS Comput. Biol.} {\bf 4}, e1000216(25).

\bibitem{yuan:2008}
Yuan, G.~C. and Liu, J.~S. (2008)
{\em PLoS Comput. Biol.} {\bf 4}, 0164--0174.

\bibitem{peckham:2007}
Peckham, H.~E., Thurman, R.~E., Fu, Y.~T., Stamatoyannopoulos, J.~A., Noble,
  W.~S., Struhl, K., and Weng, Z.~P. (2007)
{\em Genome Res.} {\bf 17}, 1170--1177.

\bibitem{lee:2007}
Lee, W., Tillo, D., Bray, N., Morse, R.~H., Davis, R.~W., Hughes, T.~R., and
  Nislow, C. (2007)
{\em Nature Genet.} {\bf 39}, 1235--1244.

\bibitem{morozov:2009}
Morozov, A.~V., Fortney, K., Gaykalova, D.~A., Studitsky, V.~M., Widom, J., and
  Siggia, E.~D. (2009)
{\em Nucleic Acids Res.} {\bf 37}, 4707--4722.

\bibitem{miele:2008}
Miele, V., Vaillant, C., d'AubentonCarafa, Y., Thermes, C., and Grange, T.
  (2008)
{\em Nucleic Acids Res.} {\bf 36}, 3746--3756.

\bibitem{tolstorukov:2007}
Tolstorukov, M.~Y., Colasanti, A.~V., McCandlish, D.~M., Olson, W.~K., and
  Zhurkin, V.~B. (2007)
{\em J. Mol. Biol.} {\bf 371}, 725--738.

\bibitem{zhang:2009}
Zhang, Y., Moqtaderi, Z., Rattner, B.~P., Euskirchen, G., Snyder, M., Kadonaga,
  J.~T., Liu, X.~S., and Struhl, K. (2009)
{\em Nature Struct. Mol. Biol.} {\bf 16}, 847--852.

\bibitem{percus:1976}
Percus, J.~K. (1976)
{\em J. Stat. Phys.} {\bf 15}, 505--511.

\bibitem{ulanovsky:1986}
Ulanovsky, L.~E. and Trifonov, E.~N.
Biomolecular Stereodynamics III pp. 35--44
Adenine Press New York (1986).

\bibitem{wang:2008}
Wang, J.~P., Fondufe-Mittendorf, Y., Xi, L., Tsai, G.~F., Segal, E., and Widom,
  J. (2008)
{\em PLoS Comput. Biol.} {\bf 4}, e1000175.

\bibitem{lubliner:2009}
Lubliner, S. and Segal, E. (2009)
{\em Bioinformatics} {\bf 25}, I348--I355.

\bibitem{tillo:2009}
Tillo, D. and Hughes, T.~R. (2009)
{\em BMC Bioinformatics} {\bf 10}, 442.

\bibitem{valouev:2008}
Valouev, A., Ichikawa, J., Tonthat, T., Stuart, J., Ranade, S., Peckham, H.,
  Zeng, K., Malek, J.~A., Costa, G., McKernan, K., Sidow, A., Fire, A., and
  Johnson, S.~M. (2008)
{\em Genome Res.} {\bf 18(7)}, 1051--1063.

\bibitem{horz:1981}
H\"{o}rz, W. and Altenburger, W. (1981)
{\em Nucleic Acids Res.} {\bf 9}, 2643--2658.

\bibitem{yuan:2005}
Yuan, G.~C., Liu, Y.~J., Dion, M.~F., Slack, M.~D., Wu, L.~F., Altschuler,
  S.~J., and Rando, O.~J. (2005)
{\em Science} {\bf 309}, 626--630.

\bibitem{zawadzki:2009}
Zawadzki, K.~A., Morozov, A.~V., and Broach, J.~R. (2009)
{\em Mol. Biol. Cell} {\bf 20}, 3503--3513.

\bibitem{widom:1992}
Widom, J. (1992)
{\em Proc. Natl. Acad. Sci. USA} {\bf 89}, 1095--1099.

\bibitem{flaus:1996}
Flaus, A., Luger, K., Tan, S., and Richmond, T. (1996)
{\em Proc. Natl. Acad. Sci. USA} {\bf 93}, 1370--1375.

\bibitem{kassabov:2002}
Kassabov, S., Henry, N., Zofall, M., Tsukiyama, T., and Bartholomew, B. (2002)
{\em Mol. Cell. Biol.} {\bf 22}, 7524--7534.

\bibitem{davey:2004}
Davey, C., Pennings, S., Reilly, C., Meehan, R., and Allan, J. (2004)
{\em Nucl. Acids Res.} {\bf 32}, 4322--4331.

\bibitem{lowary:1998}
Lowary, P. and Widom, J. (1998)
{\em J. Mol. Biol.} {\bf 276}, 19--42.

\bibitem{nagalakshmi:2008}
Nagalakshmi, U., Wang, Z., Waern, K., Shou, C., Raha, D., Gerstein, M., and
  Snyder, M. (2008)
{\em Science} {\bf 320}, 1344--1349.

\end{thebibliography}

\end{document}